\documentclass[usenatbib]{mn2e}
\usepackage{mncite}
\usepackage{graphicx}
\usepackage{hyperref}
\usepackage{amssymb}
\usepackage{amsmath}
\usepackage{aas_macros}
\usepackage{ulem}
\usepackage{longtable}
\usepackage{multicol}
\usepackage{supertabular}
\usepackage{caption}
\usepackage{subcaption}
\usepackage{color}
\usepackage{lscape}

\title[Age discrepancies for exoplanet hosts?]{Discrepancies between isochrone fitting and gyrochronology for exoplanet host stars?}
\author[D. J. A. Brown et al.]
{D. J. A. Brown$^{1,2,3}$\thanks{E-mail: d.j.a.brown@warwick.ac.uk}\\
$^{1}$ Department of Physics, University of Warwick, Gibbet Hill Road, Coventry CV4 7AL, UK.\\
$^{2}$ Astrophysics Research Centre, School of Mathematics and Physics, Queen's University, University Road, Belfast BT7 1NN, UK.\\
$^{3}$ SUPA, School of Physics and Astronomy, University of St Andrews, North Haugh, St Andrews, Fife KY16 9SS, UK.\\
}

\begin{document}

\date{Accepted 0000 December 00. Received 0000 December 00; in original form 0000 October 00}

\pagerange{\pageref{firstpage}--\pageref{lastpage}} \pubyear{2014}

\maketitle

\label{firstpage}

\begin{abstract}
Using a sample of $68$ planet-hosting stars I carry out a comparison of isochrone fitting and gyrochronology to investigate whether tidal interactions between the stars and their planets are leading to underestimated ages using the latter method. I find a slight tendency for isochrones to produce older age estimates but find no correlation with tidal time-scale, although for some individual systems the effect of tides might be leading to more rapid rotation than expected from the stars' isochronal age, and therefore an underestimated gyrochronology age. By comparing to planetary systems in stellar clusters, I also find that in some cases isochrone fitting can overestimate the age of the star. The evidence for any bias on a sample-wide level is inconclusive.

I also consider the subset of my sample for which the sky-projected alignment angle between the stellar rotation axis and the planet's orbital axis has been measured, finding similar patterns to those identified in the full sample. However, small sample sizes for both the misaligned and aligned systems prevent strong conclusions from being drawn.
\end{abstract}


\section{Introduction}
\label{sec:intro}
Determining stellar ages is notoriously difficult, but they are becoming increasingly important in the field of exoplanetary science as a stepping stone to a better understanding of the evolution of planetary systems. In order to fully characterize the time-scales involved in processes such as planet formation and destruction, orbital migration and circularization, and intra-system dynamical interactions, it is vital that we are able to accurately assess the ages of exoplanet host stars. A wide range of methods exist for the evaluation of stellar age, making use of a disparate array of phenomena. Two that are particularly prevalent in the exoplanet literature are gyrochronology and isochrone fitting.

\subsection{Isochrone fitting}
\label{sec:isochrones}
Stellar model fitting, also known as isochrone fitting, is widely used owing to its relative ease of implementation. Traditionally, either absolute stellar magnitude, $M_{\rm v}$ \citep[e.g.][]{1993AA...275..101E,1999AA...348..897L}, or stellar surface gravity, $\log(g_{\rm s})$ \citep[e.g.][]{2005AA...431.1105B,2005ApJ...624..372K}, is interpolated through theoretical models of stellar evolution along with the stellar effective temperature, $T_{\rm eff}$. However for exoplanetary studies it has become common practice to replace $M_{\rm v}$ and $\log(g_{\rm s})$ with the cube root of the stellar density, as this can be constrained to high precision through transit photometry. This leads to a parameter space of [$T_{\rm eff}, (\rho_s/\rho_\odot)^{-1/3}$] \citep{2007ApJ...664.1190S}.

In principle, isochrone fitting is applicable to stars across the spectral range, but it can be difficult to determine ages for stars with spectral type later than mid-to-late G owing to the fact that they evolve very slowly, having nuclear burning time-scales that are longer than the age of the Galactic disc. The complex shape of isochrones close to the main-sequence (MS) turn-off can also pose problems, and linearly interpolating through isochrones is not always a valid approach owing to their non-uniform spacing \citep{2010ARAA..48..581S}.

\subsection{Gyrochronology}
\label{sec:gyro}
Gyrochronology is a method for determining a cool star's age through measurement of its rotation period and colour, and arose from observations showing that by the age of the Hyades the rotation of stars in stellar clusters tends to converge to a single period--colour--age relation. First suggested by \citet{2003ApJ...586..464B}, it builds on the simple relationship between rotation period and age described by \citet{1972ApJ...171..565S} to provide a model-independent alternative to age estimation methods that require distance measurements for the stars under examination. Subsequent development of the method in \citet{2007ApJ...669.1167B} showed that gyrochronology provides age estimate that are more self-consistent than those derived through isochrone fitting. It has been demonstrated that, if rotation periods have been measured and the equations correctly calibrated, gyrochronology can provide ages with an accuracy of 10\,percent for F, G, K, and M spectral types \citep{2008ApJ...687.1264M,2009MNRAS.400..451C,2011ASPC..448..841D}. 

One drawback with the method is that it assumes that the natural rotational evolution of the star progresses free from any outside influence. This is not always the case; in both binary star systems and hot Jupiter exoplanetary systems, tidal torques between nearby bodies in close proximity can potentially overwhelm the natural spin-down that results from magnetic braking, at least for short periods of time. In addition, gyrochronology is not calibrated for hot, rapidly rotating, early-type stars, and is only limited to `solar-type (FGKM) stars'  \citep{2007ApJ...669.1167B}. As transit searches prioritize stars of F or G spectral type (e.g. \citealt[][for WASP targets]{bentley2010}; \citealt[][for \textit{Kepler} targets]{2010ApJ...713L.109B}), this is not particularly limiting, but \citet{2010AA...512A..77L} suggests that gyrochronology may not always provide accurate age estimates for planetary systems. \citeauthor{2010AA...512A..77L} found that  plotting $P_{\rm rot}t^{-\zeta}$ as a function of $T_{\rm eff}$ for planet-hosting stars gives a poor fit to the period--colour relation of \citet{2007ApJ...669.1167B}, and that the rotation periods of hot Jupiter hosts were, on average, a factor of $0.7$ faster than non-planet-hosting stars; such a discrepancy would clearly lead to underestimation of the gyrochronology ages of stars with known planets with respect to their true age.

In this work, I investigate the ages of a sample of planetary systems primarily discovered by transit searches. I first discuss the methods that I have used to determine the ages of the stars in my sample, before comparing the results obtained using isochrone fitting to those obtained through gyrochronology. I also investigate the subset of my sample for which the sky-projected spin-orbit alignment angle has been measured, to check for biases in either of the age estimation methods that might be induced in misaligned systems.

\section{Implementation}
\label{sec:age_methods}
I consider a sample of $68$ planet-hosting stars with $6226$\,K\,$\leq T_{\rm eff}\leq5273$\,K. These limits were chosen to restrict my sample to spectral types F7--G9 (inclusive), and are based on the values given in table\,B1 of \citet{2008oasp.book.....G}. This restriction on the available parameter space avoids the problems encountered when isochrone fitting for stars with long MS lifetimes, and has an upper limit that coincides with the magnetic braking boundary at mid-to-late F spectral type observed by \citet{1967ApJ...150..551K}. Stars with earlier spectral types than this show little-to-no relation between $P_{\rm rot}$ and age \citep{1986ApJ...310..360W}, and are therefore poor targets for gyrochronology.

\begin{figure}
	\centering
	\includegraphics[width=0.48\textwidth]{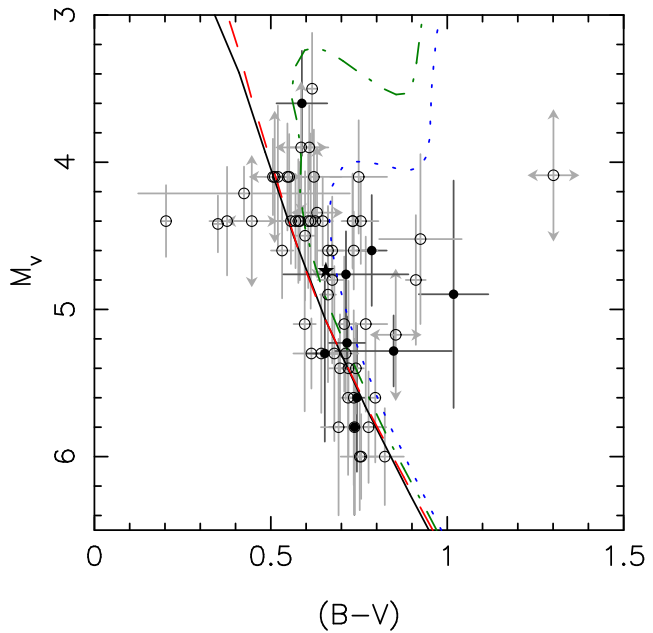}
	\caption{Colour--magnitude diagram for the sample of stars detailed in Table\,\ref{tab:data}. Solid circles represent stars with measured rotation periods and open circles represent those for which the rotation period was derived from stellar data. The solid star represents the location of the Sun. Various isochrones from the YY set are also represented: the ZAMS (black, solid line); $1$\,Gyr (red, dashed line); $5$\,Gyr (green, dot--dashed line), and $10$\,Gyr (blue, dotted line). Note that the position of the isochrones shifts slightly depending on the choice of models, and that the absolute magnitudes are calculated using estimated distances in the majority of cases.}
	\label{fig:CMD}
\end{figure}

The majority of the sample, which is described in Table\,\ref{tab:data}, consists of the host stars of sub-stellar companions discovered by the WASP project \citep{2006PASP..118.1407P}, with the remaining systems selected from the Holt--Rossiter--McLaughlin data base of Ren\'{e} Heller \footnote{www.physics.mcmaster.ca/$\sim$rheller/} as of 2013 October 24. Figure\,\ref{fig:CMD} displays a colour--magnitude diagram for the sample as compared to the Yonsei--Yale (YY) isochrones for the zero-age main sequence (ZAMS) and other representative ages. I note that there are three systems which seem to lie to the left of the ZAMS (as well as a further two for which the uncertainties are such that agreement with the ZAMS is possible) in a position which seems to be somewhat unphysical. Either the \textit{(B-V)} colours for these systems are substantially wrong or they are very young systems, although this seems unlikely given the selection constraints placed on my sample.

\onecolumn
\begin{landscape}
\begingroup
\scriptsize
\begin{longtable}{lllllllllllllll}
\caption{Data for the sample of $68$ stars for which I compare the isochronal and gyrochronological ages. Systems for which the rotation period has been directly measured have been placed at the head of the table, with the exception of Kepler-30. Systems which were disregarded during the analysis, either for having one or both of the age estimation methods return a null result, or for having one or both methods return an age greater than that currently accepted as the age of the Universe, have been separated out and moved to the foot of the table. } \\
\label{tab:data} \\
\hline
System	&	$T_{\rm eff}$	& $[M/H]$	& $M_{\rm star}$	& $R_{\rm star}$	& $\rho_{\rm star}$	& YY age	& $v\sin I$		& $i_{\rm orb}$	& $P_{\rm rot, m}$	& $P_{\rm rot, d}$	& \textit{(B-V)}	& \textit{(J-K)}	& ${\rm Gyro}$ ${\rm age}_4$	& Ref.$^{[1,2]}$ \\ [2pt]
		&	(K)			&		 	& ($M_{\odot}$)		& ($R_{\odot}$)		& ($\rho_{\odot}$)	& (Gyr)		& (km s$^{-1}$)	& ($^{\circ}$)	& (d)					& (d)				&		&		& (Gyr)	&		   \\ [2pt]
\hline
\endfirsthead
\multicolumn{15}{l}{\tablename\ \thetable\ -- \textit{Continued from previous page}} \\
\hline
System	&	$T_{\rm eff}$	& $[M/H]$	& $M_{\rm star}$	& $R_{\rm star}$	& $\rho_{\rm star}$	& YY age	& $v\sin I$		& $i_{\rm orb}$	& $P_{\rm rot, m}$	& $P_{\rm rot, d}$	& \textit{(B-V)}	& \textit{(J-K)}	& ${\rm Gyro}$ ${\rm age}_4$	& Ref.$^{[1,2]}$ \\ [2pt]
		&	(K)			&		 	& ($M_{\odot}$)		& ($R_{\odot}$)		& ($\rho_{\odot}$)	& (Gyr)		& (km s$^{-1}$)	& ($^{\circ}$)	& (d)					& (d)				&		&		& (Gyr)	&		   \\ [2pt]
\hline
\endhead
\hline
\multicolumn{15}{r}{\textit{Continued on next page}} \\
\endfoot
\hline
\endlastfoot
	WASP-4	& $5500\pm100$	& $-0.03$	& $0.930^{+0.054}_{-0.053}$	& $0.907^{+0.014}_{-0.013}$	& $1.27\pm0.01$		& $5.13^{+1.98}_{-1.76}$	& $2.14^{+0.38}_{-0.35}$ & $88.8^{+0.6}_{-0.4}$ 	& $22.2^{+3.3}_{-3.3}$	& $21.4^{5.0}_{-3.4}$	& $0.744\pm0.022$	& $0.433\pm0.033$		& $3.02^{+0.83}_{-0.35}$ & a; A \\ [2pt]
	WASP-19	& $5475\pm98$	& $0.02$	& $0.969^{+0.023}_{-0.023}$	& $0.993^{+0.018}_{-0.018}$	& $0.990\pm0.043$		& $8.91^{+2.21}_{-0.92}$	& $4.63^{+0.27}_{-0.27}$	& $79.42^{+0.39}_{-0.39}$ & $10.5\pm0.2$		& $5.7^{+0.3}_{-0.3}$	& $0.737\pm0.072$& $0.430\pm0.035$		& $0.74^{+0.05}_{-0.04}$	& b; B, C \\ [2pt]
	WASP-46	& $5622\pm135$	& $-0.37$	& $0.956^{+0.034}_{-0.034}$	& $0.917^{+0.028}_{-0.028}$	& $1.20\pm0.12$		& $10.84^{+3.81}_{-4.03}$& $1.9^{+1.2}_{-1.2}$	& $82.63^{+0.38}_{-0.38}$ & $16.1^{+1.0}_{-1.0}$ 		& $23.3^{+27.7}_{-8.6}$	& $0.653\pm0.051$	& $0.352\pm0.035$		& $1.74^{+0.30}_{-0.24}$	& c; D, E \\ [2pt]
	WASP-50	& $5857\pm133$	& $-0.12$	& $0.861^{+0.052}_{-0.052}$	& $0.855^{+0.018}_{-0.018}$	& $1.376\pm0.032$		& $1.86^{+4.41}_{-1.20}$	& $2.6^{+0.5}_{-0.5}$	& $84.74^{+0.24}_{-0.24}$ & $16.3^{+0.5}_{-0.5}$ 		& $16.6^{+4.0}_{-2.7}$	& $0.786\pm0.042$	& $0.432\pm0.03$		& $2.23^{+0.50}_{-0.31}$	& d; D, F, G \\ [2pt]
	CoRoT-2	& $5575\pm66$	& $-0.04$	& $0.97^{+0.06}_{-0.06}$		& $0.902^{+0.018}_{-0.018}$	& $1.32\pm0.11$		& $3.01^{+2.26}_{-1.50}$	& $11.95^{+0.58}_{-0.55}$ & $87.84^{+0.16}_{-0.17}$ & $4.5^{+0.14}_{-0.14}$	& $12.5^{+3.1}_{-2.0}$	& $1.018\pm0.098$ & $0.473\pm0.041$	& $0.20^{+0.01}_{-0.01}$	& e; H \\ [2pt]
	CoRoT-18 & $5440\pm100$	& $-0.1$	& $0.95^{+0.15}_{-0.15}$		& $1.00^{+0.13}_{-0.13}$		& $0.95\pm0.40$		& $11.80^{+5.71}_{-9.80}$& $8.0^{+1.0}_{-1.0}$	& $86.5^{+1.4}_{-0.9}$	& $5.4^{+0.4}_{-0.4}$		& $6.3^{+1.3}_{-1.1}$	& $0.848\pm0.165$ & $0.427\pm0.038$	& $0.25^{+0.03}_{-0.03}$	& f; I \\ [2pt]
	Kepler-17	& $5630\pm100$	& $0.3$	& $1.06^{+0.07}_{-0.07}$		& $1.02^{+0.03}_{-0.03}$		& $1.00\pm0.11$		& $2.21^{+2.00}_{-1.17}$	& $4.7^{+1.0}_{-1.0}$	& $87.2^{+0.15}_{-0.15}$	& $11.89^{+0.15}_{-0.15}$	& $11.0^{+3.0}_{-2.0}$	& $0.713\pm0.177$ & $0.407\pm0.031$	& $1.00^{+0.08}_{-0.07}$	& g \\ [2pt]
	Kepler-63  & $5576\pm50$	& $0.05$	& $0.98\pm0.04$			& $0.901^{+0.027}_{-0.022}$	& $1.345^{+0.089}_{-0.083}$	& $1.77^{+1.25}_{-1.41}$	& $5.6\pm0.8$		& $87.81\pm0.02$		& $5.401\pm0.014$			& $8.17^{+1.37}_{-1.02}$	& $0.716\pm0.051$	& $0.402\pm0.032$		& $0.26^{+0.004}_{-0.003}$ & h; HHH \\ [2pt]
	\hline \\
	WASP-1	& $6111\pm44$	& $0.26$	& $1.208^{+0.012}_{- 0.012}$	& $1.462^{+0.019}_{-0.019}$	& $0.39\pm0.01$		& $2.71^{+0.21}_{-0.17}$	& $5.79\pm0.35$		& $89.2\pm0.8$		& $-$	& $12.8^{+0.8}_{-0.7}$	& $0.617\pm0.014$& $0.310\pm0.026$		& $2.41^{+0.52}_{-0.38}$	& J, K \\ [2pt]
	WASP-5	& $5700\pm100$	& $0.09$	& $1.0^{+0.063}_{-0.064}$	& $1.077^{+0.042}_{-0.042}$	& $0.84^{+0.06}_{-0.14}$	& $5.04^{+2.65}_{-1.62}$	& $3.24^{+0.35}_{-0.27}$ & $86.1^{+0.7}_{-1.5}$ 	& $-$	& $16.7^{+1.7}_{-1.7}$ 	& $0.662\pm0.022$ & $0.351\pm0.032$	& $2.00^{+0.46}_{-0.39}$	& A \\ [2pt]
	WASP-6	& $5450\pm100$	& $-0.20$	& $0.888^{+0.050}_{-0.080}$	& $0.870^{+0.025}_{-0.036}$	& $1.34\pm0.11$		& $8.45^{+3.25}_{-3.29}$	& $1.6^{+0.27}_{-0.17}$ & $88.47^{+0.65}_{-0.47}$& $-$	& $27.1^{+3.6}_{-3.8}$ 	& $0.796\pm0.014$ & $0.444\pm0.036$	& $4.10^{+1.20}_{-1.03}$	& A \\ [2pt]
	WASP-8	& $5600\pm80$	& $0.17$	& $1.030^{+0.054}_{-0.06}$	& $0.945^{+0.051}_{-0.036}$	& $1.22^{+0.17}_{-0.15}$	& $<3.58$				& $1.59^{+0.08}_{-0.09}$	& $88.55^{+0.15}_{-0.17}$ & $-$	& $30.3^{+2.3}_{-2.0}$	& $0.615\pm0.051$	 & $0.415\pm0.035$	& $5.72^{+0.98}_{-0.78}$	& L \\ [2pt]
	WASP-12	& $6118\pm64$	& $0.07$	& $1.35^{+0.14}_{-0.14}$		& $1.63^{+0.08}_{-0.08}$		& $0.315\pm0.09$		& $3.49^{+1.32}_{-0.26}$	& $3.4^{+0.9}_{-0.9}$	& $82.5^{+0.8}_{-0.7}$	& $-$	& $24.1^{+8.7}_{-5.3}$	& $0.578\pm0.073$	 & $0.289\pm0.029$	& $8.10^{+7.26}_{-3.31}$	& M, N, O \\ [2pt]
	WASP-16	& $5706\pm155$	& $0.0$	& $1.01^{+ 0.05}_{-0.06}$		& $0.983^{+0.047}_{-0.049}$	& $1.07^{+0.14}_{-0.12}$ & $3.37^{+3.36}_{-2.17}$	& $1.47^{+0.30}_{-0.32}$ & $84.86^{+0.32}_{-0.32}$ & $-$	& $33.8^{+8.9}_{-6.1}$	& $0.696\pm0.032$& $0.395\pm0.033$		& $7.81^{+5.29}_{-2.58}$	& P \\ [2pt]
	WASP-21	& $5800\pm100$	& $-0.4$	& $1.02^{+0.05}_{-0.05}$		& $1.06^{+0.04}_{-0.04}$		& $0.65^{+0.04}_{-0.06}$ & $12.37^{+2.77}_{-1.90}$& $1.5^{+0.6}_{-0.6}$	& $88.75^{+0.84}_{-0.70}$ & $-$	& $35.7^{+23.6}_{-10.1}$& $0.532\pm0.032$& $0.368\pm0.029$	 & $9.80^{+16.59}_{-4.85}$	& Q, R \\ [2pt]
	WASP-22	& $5958\pm98$	& $0.05$	& $1.109^{+0.026}_{-0.026}$	& $1.219^{+0.052}_{-0.033}$	& $0.61^{+0.05}_{-0.07}$ & $4.25^{+1.17}_{-1.01}$	& $4.42^{+0.34}_{-0.34}$	& $88.26^{+0.91}_{-0.91}$ & $-$	& $14.0^{+1.3}_{-1.1}$	& $0.597\pm0.028$	& $0.316\pm0.029$		& $1.94^{+0.57}_{-0.38}$	& S \\ [2pt]
	WASP-25	& $5785\pm94$	& $-0.11$	& $0.95^{+ 0.04}_{-0.04}$		& $0.910^{+0.028}_{-0.029}$	& $1.26^{+0.09}_{-0.08}$ & $1.94^{+1.75}_{-1.79}$	& $2.83^{+0.26}_{-0.27}$ & $87.83^{+0.31}_{-0.27}$& $-$	& $16.2^{+1.8}_{-1.5}$ 	& $0.708\pm0.022$& $0.422\pm0.034$		& $2.07^{+0.51}_{-0.39}$	& P \\ [2pt]
	WASP-26	& $5939\pm100$	& $-0.02$	& $1.111^{+0.028}_{-0.028}$	& $1.303^{+0.059}_{-0.059}$	& $0.502\pm0.062$		& $5.73^{+1.50}_{-1.41}$	& $2.2^{+0.7}_{-0.7}$	& $82.91^{+0.46}_{-0.46}$ & $-$	& $29.6^{+13.7}_{-7.3}$	& $0.626\pm0.050$	& $0.331\pm0.032$		& $8.17^{+9.78}_{-3.60}$	& S \\ [2pt]
	WASP-28	& $6175\pm142$	& $-0.29$	& $1.08^{+0.04}_{-0.04}$		& $1.05^{+0.06}_{-0.06}$		& $0.93\pm0.13$		& $1.68^{+2.65}_{-0.96}$	& $4.1^{+0.6}_{-0.6}$	& $89.1^{+0.6}_{-0.6}$	& $-$	& $13.0^{+2.3}_{-1.8}$	& $0.582\pm0.014$	& $0.346\pm0.035$		& $2.83^{+2.53}_{-1.13}$	& T \\ [2pt]
	WASP-30	& $6202^{+42}_{-51}$	& $0.083$& $1.249^{+0.032}_{-0.036}$	& $1.389^{+0.033}_{-0.025}$	& $0.47^{+0.02}_{-0.03}$	& $3.52^{+0.32}_{-0.60}$& $12.1^{+0.4}_{-0.5}$	& $89.43^{+0.51}_{-0.93}$ & $-$	& $5.8^{+0.3}_{-0.2}$ & $0.520\pm0.014$ & $0.309\pm0.035$	& $0.62^{+0.13}_{-0.10}$	& U \\ [2pt]
	WASP-32	& $6077\pm99$	& $-0.13$	& $1.07^{+0.05}_{-0.05}$		& $1.087^{+0.031}_{-0.032}$	& $0.84\pm0.05$		& $2.10^{+1.54}_{-1.35}$	& $3.94^{+0.42}_{-0.48}$	& $85.08^{+0.24}_{-0.22}$ & $-$	& $13.9^{+1.9}_{-1.4}$ 	& $0.588\pm0.072$ & $0.342\pm0.032$		& $2.47^{+1.14}_{-0.66}$	& V \\ [2pt]
	WASP-35	& $6001\pm74$	& $-0.15$	& $1.07^{+0.08}_{-0.08}$		& $1.09^{+0.14}_{-0.14}$		& $0.83\pm0.07$		& $2.98^{+2.16}_{-1.75}$	& $3.9^{+0.4}_{-0.4}$	& $87.96^{+0.62}_{-0.49}$ & $-$	& $14.1^{+2.4}_{-2.2}$	& $0.570\pm0.050$& $0.362\pm0.037$		& $2.13^{+0.88}_{-0.64}$	& W \\ [2pt]
	WASP-36	& $5959\pm134$	& $-0.26$	& $1.040^{+0.031}_{-0.031}$	& $0.951^{+0.018}_{-0.018}$	& $1.21\pm0.05$		& $1.86^{+1.96}_{-1.24}$	& $3.3^{+1.2}_{-1.2}$	& $83.61^{+0.21}_{-0.21}$ & $-$	& $14.5^{+8.4}_{-3.9}$	& $0.613\pm0.036$	& $0.315\pm0.038$		& $2.17^{+3.36}_{-1.04}$	& X \\ [2pt]
	WASP-37	& $5800\pm150$	& $-0.40$	& $0.925^{+0.120}_{-0.120}$	& $1.003^{+0.053}_{-0.053}$	& $0.93^{+0.06}_{-0.10}$	& $10.43^{+3.66}_{-3.30}$& $2.4^{+1.6}_{-1.6}$	& $88.82^{+0.77}_{-0.86}$ & $-$	& $19.9^{+23.9}_{-7.6}$ 	& $0.643\pm0.022$& $0.406\pm0.032$		& $3.16^{+11.80}_{-1.90}$	& Y \\ [2pt]
	WASP-38	& $6186\pm79$	& $-0.02$	& $1.22^{+0.04}_{-0.04}$		& $1.351^{+0.022}_{-0.018}$	& $0.50\pm0.01$		& $3.29^{+0.42}_{-0.53}$	& $7.49^{+0.15}_{-0.16}$	& $89.46^{+0.32}_{-0.37}$ & $-$	& $9.1^{+0.2}_{-0.2}$& $\sim0.511$& $0.289\pm0.046$	& $1.43^{+0.54}_{-0.31}$	& V \\ [2pt]
	WASP-39	& $5406\pm143$	& $-0.12$	& $0.93^{+0.03}_{-0.03}$		& $0.895^{+0.023}_{-0.023}$	& $1.30^{+0.08}_{-0.07}$ & $8.55^{+1.99}_{-4.02}$	& $1.4^{+0.6}_{-0.6}$	& $87.83^{+0.25}_{-0.22}$ & $-$	& $32.0^{+22.7}_{-9.5}$	& $0.777\pm0.050$	& $0.461\pm0.033$		& $5.51^{+10.20}_{-2.72}$	& Z \\ [2pt]
	WASP-41	& $5450\pm150$	& $-0.08$	& $0.93^{+0.03}_{-0.03}$		& $0.90^{+0.05}_{-0.05}$		& $1.27\pm0.14$		& $6.97^{+4.57}_{-3.34}$	& $1.6^{+1.1}_{-1.1}$	& $87.7^{+0.08}_{-0.08}$	& $-$	& $26.9^{+35.0}_{-10.6}$	& $0.752\pm0.054$	& $0.419\pm0.031$		& $4.07^{+16.48}_{-2.49}$	& AA \\ [2pt]
	WASP-47	& $5402\pm115$	& $0.18$	& $1.084^{+0.037}_{-0.037}$	& $1.15^{+0.03}_{-0.02}$		& $0.71^{+0.02}_{-0.04}$ & $11.28^{+2.94}_{-2.35}$& $3.0^{+0.6}_{-0.6}$	& $89.2^{+0.5}_{-0.7}$	& $-$	& $19.4^{+4.9}_{-3.2}$	& $0.735\pm0.014$	& $0.421\pm0.034$		& $2.13^{+1.15}_{-0.61}$	& D, BB \\ [2pt]
	WASP-48	& $6000\pm138$	& $-0.12$	& $1.1^{+ 0.05}_{-0.05}$		& $1.09^{+0.14}_{-0.14}$		& $0.22\pm0.03$		& $5.39^{+0.63}_{-1.77}$	& $12.2^{+0.7}_{-0.7}$	& $80.09^{+0.88}_{-0.79}$ & $-$	& $4.5^{+0.7}_{-0.6}$ & $0.749\pm0.081$& $0.255\pm0.033$	& $0.24^{+0.10}_{-0.06}$	& W \\ [2pt]
	WASP-54	& $6100\pm100$	& $-0.27$	& $1.213^{+0.032}_{-0.032}$	& $1.828^{+0.091}_{-0.081}$	& $0.21^{+0.06}_{-0.02}$	& $5.56^{+0.89}_{-0.51}$	& $4.0^{+0.8}_{-0.8}$	& $84.97^{+0.63}_{-0.59}$ & $-$	& $23.1^{+5.9}_{-4.0}$	& $0.557\pm0.036$ & $0.330\pm0.032$	& $6.47^{+5.84}_{-2.36}$	& CC \\ [2pt]
	WASP-55	& $5947\pm129$	& $-0.20$	& $1.01^{+0.04}_{-0.04}$		& $1.06^{+0.03}_{-0.02}$		& $0.85^{+0.03}_{-0.07}$	& $5.33^{+2.17}_{-2.35}$	& $3.1^{+1.0}_{-1.0}$	& $89.2^{+0.6}_{-0.6}$	& $-$	& $17.3^{+8.5}_{-4.2}$	& $0.606\pm0.063$	 & $0.379\pm0.036$	& $2.97^{+3.66}_{-1.34}$	& D, BB \\ [2pt]
	WASP-57	& $5600\pm100$	& $-0.25$	& $0.954^{+0.027}_{-0.027}$	& $0.836^{+0.07}_{-0.16}$	& $1.638^{+0.044}_{-0.063}$ & $2.12^{+1.81}_{-1.81}$& $3.7^{+1.3}_{-1.3}$	& $88.0^{+0.1}_{-0.2}$	& $-$	& $11.0^{+6.2}_{-3.2}$ 	& $0.719\pm0.022$ & $0.381\pm0.035$	&  $0.86^{+1.08}_{-0.38}$	& CC \\ [2pt]
	WASP-58	& $5900\pm100$	& $-0.46$	& $0.94^{+0.10}_{-0.10}$		& $1.25^{+0.17}_{-0.17}$		& $0.64\pm0.16$		& $9.75^{+3.90}_{-4.66}$	& $2.8^{+0.9}_{-0.9}$	& $86.97^{+1.55}_{-1.55}$ & $-$	& $22.6^{+11.7}_{-6.1}$ 	& $0.376\pm0.050$ & $0.341\pm0.036$	& $4.54^{+5.97}_{-2.11}$	& DD \\ [2pt]
	WASP-60	& $5900\pm100$	& $-0.04$& $1.078^{+0.035}_{-0.035}$	& $1.14^{+0.13}_{-0.13}$		& $0.72\pm0.20$		& $3.51^{+2.68}_{-1.45}$	& $3.4^{+0.8}_{-0.8}$	& $87.86^{+1.61}_{-1.61}$ & $-$	& $16.8^{+5.6}_{-3.6}$ 	& $0.680\pm0.014$ & $0.379\pm0.031$	& $2.54^{+2.04}_{-0.98}$	& DD \\ [2pt]
	WASP-64	& $5635\pm143$	& $-0.08$	& $1.004^{+0.028}_{-0.028}$	& $1.058^{+0.025}_{-0.025}$	& $0.85^{+0.05}_{-0.04}$	& $8.94^{+3.15}_{-2.55}$	& $3.4^{+0.8}_{-0.8}$	& $86.57^{+0.80}_{-0.60}$ & $-$	& $15.7^{+4.9}_{-3.1}$	& $0.720\pm0.028$	 & $0.412\pm0.029$	& $1.71^{+1.18}_{-0.59}$	& D, EE \\ [2pt]
	WASP-65	& $5600\pm100$	& $-0.06$	& $1.00^{+0.02}_{-0.02}$		& $1.07^{+0.01}_{-0.01}$		& $0.91\pm0.04$		& $8.92^{+1.87}_{-1.97}$	& $3.6^{+0.5}_{-0.5}$	& $87.45^{+0.15}_{-0.13}$ & $-$	& $15.0^{+2.4}_{-1.8}$	& $0.674\pm0.041$ & $0.323\pm0.030$	& $1.51^{+0.50}_{-0.33}$	& FF \\ [2pt]
	WASP-70\,A	& $5763\pm79$ & $-0.006$& $1.106^{+0.042}_{-0.042}$ & $1.215^{+0.064}_{-0.089}$	& $0.62^{+0.14}_{-0.08}$	& $4.68^{+3.47}_{-1.31}$	& $1.8^{+0.4}_{-0.4}$	& $87.12^{+1.24}_{-0.65}$ & $-$	& $33.8^{+10.0}_{-6.4}$ & $0.732\pm0.032$ & $0.416\pm0.046$	& $8.29^{+5.62}_{-2.89}$ & D, XX \\ [2pt]
	WASP-71	& $6050\pm100$	& $0.15$	& $1.572^{+0.062}_{-0.062}$	& $2.32^{+0.14}_{-0.14}$		& $0.127\pm0.021$		& $3.21^{+0.38}_{-0.74}$	& $9.91^{+0.49}_{-0.49}$	& $84.2^{1.8}_{-1.8}$	& $-$	& $11.8^{+1.0}_{-0.9}$	& $0.622\pm0.078$ & $0.316\pm0.032$	& $1.64^{+0.58}_{-0.35}$	& GG \\ [2pt]
	WASP-75	& $6100\pm100$	& $0.06$	& $1.14^{+0.03}_{-0.03}$		& $1.256^{+0.029}_{-0.029}$	& $0.60\pm0.05$		& $2.08^{+0.60}_{-0.95}$	& $4.3^{+0.8}_{-0.8}$	& $82.15^{+0.21}_{-0.23}$ & $-$	& $14.6^{+3.4}_{-2.3}$	& $0.596\pm0.032$ & $0.300\pm0.035$	& $5.92^{+4.43}_{-1.78}$	& FF \\ [2pt]
	WASP-77\,A	& $5458\pm128$ & $0.11$& $0.968^{+0.030}_{-0.030}$ & $0.946^{+0.011}_{-0.010}$	& $1.14\pm0.02$		& $5.34^{+2.19}_{-2.08}$	& $4.0^{+0.2}_{-0.2}$	& $89.23^{+0.518}_{-0.670}$ & $-$	& $12.0^{+0.6}_{-0.6}$& $0.756\pm0.022$ & $0.361\pm0.037$	& $0.92^{+0.12}_{-0.09}$	& D \\ [2pt]
	WASP-95	& $5830\pm140$	& $0.14$	& $1.11^{+0.09}_{-0.09}$		& $1.13^{+0.08}_{-0.04}$		& $0.78^{+0.04}_{-0.13}$ & $2.56^{+2.18}_{-0.68}$	& $3.1\pm0.6$		& $88.4^{+1.2}_{-2.1}$ & $-$	& $18.7^{+4.6}_{-3.1}$& $0.735\pm0.067$ & $0.372\pm0.038$	& $2.90^{+1.76}_{-0.96}$	& D, HH \\ [2pt]
	WASP-96	& $5500\pm150$	& $0.14$	& $1.06^{+0.09}_{-0.09}$		& $1.05^{+0.05}_{-0.05}$		& $0.922\pm0.073$		& $5.17^{+4.32}_{-1.10}$& $1.5^{+1.3}_{-1.3}$	& $85.6^{+0.2}_{-0.2}$ & $-$	& $31.6^{+50.9}_{-13.5}$ & $0.737\pm0.014$ & $0.353\pm0.035$	& $5.59^{+28.77}_{-3.56}$	& D, HH \\ [2pt]
	WASP-99 & $6150\pm100$	& $0.21$	& $1.48\pm0.10$ 			& $1.76^{+0.11}_{-0.06}$		& $0.27^{+0.02}_{-0.04}$	& $2.45^{+0.76}_{-0.30}$	& $6.8\pm0.5$	& $88.8\pm1.1$	& $-$	& $13.2^{+1.3}_{-1.1}$	& $0.203\pm0.014$		& $0.348\pm0.033$		& $2.68^{+1.32}_{-0.73}$	& HH \\ [2pt]
	CoRoT-19 & $6090\pm70$	& $-0.02$	& $1.21^{+0.05}_{-0.05}$		& $1.65^{+0.04}_{-0.04}$		& $0.269\pm0.023$		& $4.66^{+0.04}_{-1.02}$	& $6^{+1}_{-1}$		& $88.0^{+0.7}_{-0.7}$	& $-$	& $13.9^{+2.8}_{-2.0}$	& $0.924\pm0.117$ & $0.487\pm0.034$	& $2.53^{+1.27}_{-0.75}$	& II \\ [2pt]
	HAT-P-1	& $5975\pm45$	& $0.13$	& $1.133^{+0.077}_{-0.077}$	& $1.115^{+0.050}_{-0.050}$	& $0.82\pm0.12$		& $2.15^{+1.07}_{-1.18}$	& $3.75^{+0.58}_{-0.58}$	& $86.28^{+0.20}_{-0.20}$ & $-$	& $15.0^{+3.0}_{-2.1}$	& $\sim0.586$ & $0.298\pm0.028$ 		& $2.26^{+0.94}_{-0.59}$	& JJ \\ [2pt]
	HAT-P-4	& $5890\pm67$	& $0.2$	& $1.26^{+0.10}_{-0.10}$		& $1.617^{+0.057}_{-0.050}$	& $0.30\pm0.04$ 		& $3.98^{+1.72}_{-0.28}$	& $5.83^{+0.35}_{-0.35}$	& $88.76^{+0.89}_{-1.38}$ & $-$	& $14.0^{+1.0}_{-0.9}$	& $0.647\pm0.022$ & $0.330\pm0.024$ 	& $1.75^{+0.31}_{-0.25}$	& YY \\ [2pt]
	HAT-P-8	& $6223\pm67$	& $-0.04$	& $1.192^{+0.061}_{-0.043}$	& $1.475^{+0.032}_{-0.032}$	& $0.37^{+0.01}_{-0.02}$	& $3.70^{+0.39}_{-0.49}$	& $12.6^{+1.0}_{-1.0}$	& $87.5^{+1.9}_{-0.9}$	& $-$	& $5.9^{+0.5}_{-0.4}$	& $0.506\pm0.022$ & $0.261\pm0.026$		& $0.70^{+0.28}_{-0.17}$	& O, KK \\ [2pt]
	HAT-P-13	& $5640\pm90$	& $0.46$	& $1.22^{+0.05}_{-0.10}$		& $1.559^{+0.08}_{-0.08}$	& $0.32^{+0.05}_{-0.06}$	& $5.83^{+0.51}_{-2.00}$	& $1.66\pm0.37$	& $83.40\pm0.68$ & $-$	& $47.4^{+14.1}_{-8.9}$& $0.755\pm0.05$ & $0.353\pm0.025$		& $14.17^{+9.84}_{-4.78}$	& O, TT \\ [2pt]
	HAT-P-16	& $6158\pm80$	& $0.12$	& $1.218^{+0.039}_{-0.039}$	& $1.237^{+0.054}_{-0.054}$	& $0.643\pm0.087$		& $1.97^{+0.89}_{-0.79}$	& $3.9^{+0.8}_{-0.8}$	& $86.6^{+0.7}_{-0.7}$	& $-$	& $16.0^{+4.1}_{-2.8}$	& $0.552\pm0.036$ & $0.297\pm0.030$		& $4.08^{+3.07}_{-1.50}$	& LL, MM \\ [2pt]
	HAT-P-23	& $5905\pm80$	& $0.13$	& $1.13^{+0.035}_{-0.035}$	& $1.203^{+0.035}_{-0.035}$	& $0.649\pm0.121$		& $3.96^{+0.61}_{-1.41}$	& $7.8^{+1.6}_{-1.6}$	& $85.1^{+1.5}_{-1.5}$	& $-$	& $7.8^{+2.1}_{-1.3}$	& $\sim1.301$ & $0.312\pm0.030$		& $0.59^{+0.32}_{-0.17}$	& MM, NN \\ [2pt]
	HAT-P-32	& $6207\pm88$	& $-0.04$	& $1.160^{+0.041}_{-0.041}$	& $1.219^{+0.016}_{-0.016}$	& $0.781\pm0.041$		& $1.45^{+0.89}_{-0.55}$	& $20.6^{+1.5}_{-1.5}$	& $88.9^{+0.4}_{-0.4}$	& $-$	& $3.0^{+0.2}_{-0.2}$	& $0.547\pm0.054$ & $0.261\pm0.031$		& $0.16^{+0.07}_{-0.04}$	& OO, PP \\ [2pt]
	HD\,149026 & $6160\pm50$	& $0.24$	& $1.34^{+0.02}_{-0.020}$	& $1.534^{+0.049}_{-0.047}$	& $0.371\pm0.036$		& $2.61^{+0.20}_{-0.21}$	& $7.7^{+0.8}_{-0.8}$	& $84.5^{+0.60}_{-0.52}$ & $-$	& $10.0^{+1.2}_{-1.0}$ 	& $0.350\pm0.014$ & $-$ 	& $1.61^{+0.53}_{-0.37}$	& PP, ZZ \\ [2pt]
	HD\,17156	& $6080\pm80$	& $0.13$	& $1.24^{+0.03}_{-0.03}$		& $1.44^{+0.08}_{-0.08}$		& $0.415\pm0.070$		& $3.37^{+0.88}_{-0.44}$	& $4.18^{+0.31}_{-0.31}$	& $87.21^{+0.31}_{-0.31}$ & $-$	& $17.4^{+1.7}_{-1.5}$ 	& $0.424\pm0.300$ & $-$ 	& $3.79^{+1.28}_{-0.87}$	& AAA, BBB \\ [2pt]
	HD\,209458 & $6070\pm50$	& $0.02$	& $1.148^{+0.033}_{-0.022}$	& $1.162^{+0.012}_{-0.012}$	& $0.733\pm0.008$		& $2.27^{+0.45}_{-0.56}$	& $4.4^{+0.2}_{-0.2}$	& $86.55^{+0.03}_{-0.03}$ & $-$	& $13.3^{+0.6}_{-0.6}$	& $\sim0.631$ & $0.283\pm0.033$		& $2.17^{+0.37}_{-0.29}$	& QQ \\ [2pt]
	HD\,80606	& $5570\pm44$	& $0.26$	& $1.01^{+0.05}_{-0.05}$		& $1.007^{+0.024}_{-0.024}$	& $0.989\pm0.086$		& $3.68^{+1.55}_{-1.25}$	& $1.7^{+0.3}_{-0.3}$	& $89.27^{+0.018}_{-0.018}$ & $-$	& $29.9^{+6.5}_{-4.6}$& $\sim0.854$ & $-$ 	& $5.43^{+2.53}_{-1.47}$	& CCC \\ [2pt]
	KOI-94	& $6182\pm58$	& $0.02$	& $1.277^{+0.05}_{-0.05}$	& $1.52^{+0.14}_{-0.14}$		& $0.364\pm0.10$		& $3.20^{+0.20}_{-1.66}$	& $7.3\pm0.5$	& $89.360.07$ & $-$	& $10.5^{+1.3}_{-1.2}$	& $0.680\pm0.054$ & $0.292\pm0.029$	& $1.89^{+0.71}_{-0.48}$	& RR, SS \\ [2pt]
	TrES-4	& $6200\pm75$	& $0.14$	& $1.388^{+0.042}_{-0.042}$	& $1.798^{+0.052}_{-0.052}$	& $0.239\pm0.022$		& $2.83^{+0.64}_{-0.13}$	& $8.5^{+1.2}_{-1.2}$	& $82.81^{+0.37}_{-0.37}$ & $-$	& $10.7^{+1.7}_{-1.3}$	& $\sim0.446$ & $0.253\pm0.028$ 	& $2.11^{+1.12}_{-0.66}$	& GGG \\ [2pt]
	\hline \\
	WASP-20	& $6007\pm100$	& $-0.014$& $1.076^{+0.023}_{-0.023}$	& $0.951^{+0.29}_{-0.29}$	& $1.25\pm0.11$		& $-$				& $5.81^{+1.14}_{-0.83}$	& $89.35^{+0.54}_{-0.54}$ & $-$	& $8.1^{+3.0}_{-2.7}$& $0.609\pm0.054$& $0.311\pm0.032$	& $0.75^{+0.68}_{-0.40}$	& T \\ [2pt]
	WASP-34	& $5700\pm100$	& $-0.02$	& $1.01^{+0.07}_{-0.07}$		& $0.93^{+0.12}_{-0.12}$		& $1.26\pm0.49$		& $-$				& $1.4^{+0.6}_{-0.6}$	& $85.2^{+0.2}_{-0.2}$	& $-$	& $33.2^{+23.4}_{-10.5}$	& $0.662\pm0.028$	& $0.380\pm0.033$		& $7.72^{+15.79}_{-4.12}$	& UU \\ [2pt]
	WASP-44	& $5668\pm129$	& $0.06$	& $0.917^{+0.077}_{-0.077}$	& $0.865^{+0.025}_{-0.025}$	& $1.414\pm0.058$		& $-$				& $3.2^{+0.9}_{-0.9}$	& $86.02^{+1.11}_{-0.86}$ & $-$	& $13.7^{+5.4}_{-3.0}$	& $0.769\pm0.061$	& $0.361\pm0.035$		& $1.36^{+1.20}_{-0.51}$	& D, E, VV \\ [2pt]
	WASP-45	& $5782\pm130$	& $0.36$	& $0.909^{+0.060}_{-0.060}$	& $0.945^{+0.087}_{-0.071}$	& $1.08^{+0.27}_{-0.24}$	& $-$				& $2.3^{+0.7}_{-0.7}$	& $84.47^{+0.54}_{-0.79}$ & $-$	& $20.8^{+9.2}_{-5.1}$	& $0.911\pm0.028$	& $0.459\pm0.032$		& $3.36^{+3.59}_{-1.45}$	& D, E \\ [2pt]
	WASP-49	& $5811\pm145$	& $-0.23$	& $0.938^{+0.080}_{-0.076}$	& $0.976^{+0.034}_{-0.034}$	& $1.01\pm0.06$		& $6.23^{+2.83}_{-2.33}$	& $0.9^{+0.3}_{-0.3}$	& $84.89^{+0.19}_{-0.19}$ & $-$	& $54.7^{+27.0}_{-13.8}$& $0.712\pm0.036$ & $0.397\pm0.032$	& $23.06^{+31.23}_{-10.51}$	& D, WW \\ [2pt]
	WASP-63	& $5572\pm100$	& $0.08$	& $1.32^{+0.05}_{-0.05}$		& $1.88^{+0.10}_{-0.06}$		& $0.20^{+0.02}_{-0.03}$	& $7.82^{+1.09}_{-1.13}$	& $2.8^{+0.5}_{-0.5}$	& $87.8^{+1.3}_{-1.3}$	& $-$	& $34.1^{+7.6}_{-5.4}$	& $0.741\pm0.022$	 & $0.425\pm0.032$	& $15.78^{+7.76}_{-4.54}$	& D, BB \\ [2pt]
	WASP-84	& $5314\pm88$	& $0.0$	& $0.842^{+0.037}_{-0.037}$	& $0.748^{+0.015}_{-0.015}$	& $2.015\pm0.070$		& $-$				& $4.1^{+0.3}_{-0.3}$	& $88.37\pm0.05$ & $-$	& $9.2^{+0.7}_{-0.7}$ & $0.823\pm0.054$ & $0.491\pm0.035$ 	& $0.56^{+0.08}_{-0.06}$	& D, XX \\ [2pt]
	WASP-97	& $5670\pm110$	& $0.23$	& $1.12^{+0.06}_{-0.06}$		& $1.06^{+0.04}_{-0.04}$		& $0.93\pm0.09$		& $3.21^{+1.40}_{-1.41}$	& $1.1\pm0.5$ 		& $88.0^{+1.3}_{-1.1}$ & $-$	& $48.7^{+39.5}_{-15.0}$& $0.674\pm0.032$ & $0.377\pm0.037$	& $15.53^{+32.60}_{-8.09}$	& D, HH \\ [2pt]
	WASP-98	& $5550\pm140$	& $-0.60$	& $0.69^{+0.06}_{-0.06}$		& $0.70^{+0.02}_{-0.02}$		& $1.99\pm0.07$		& $6.71^{+5.43}_{-3.66}$				& $<0.5$		& $86.3^{+0.1}_{-0.1}$ & $-$	& $70.7^{+2.0}_{-2.0}$& $0.692\pm0.050$ & $0.407\pm0.035$	& $28.84^{+4.40}_{-3.38}$	& HH, D \\ [2pt]
	Kepler-30	& $5498\pm54$	& $0.18$	& $0.99^{+0.08}_{-0.08}$		& $0.95^{+0.12}_{-0.12}$		& $1.418\pm0.071$		& $-$				& $1.94^{+0.22}_{-0.22}$	& $89.82^{+0.17}_{-0.17}$ & $16.0^{+0.4}_{-0.4}$	& $24.8^{+4.6}_{-3.9}$	& $-$ & $0.416\pm0.057$	& $1.57^{+0.09}_{-0.09}$	& i; DDD \\ [2pt]
	TrES-2	& $5850\pm50$	& $-0.01$	& $0.98^{+0.062}_{-0.062}$	& $1.00^{+0.036}_{-0.036}$	& $0.98\pm0.12$		& $3.15^{+1.40}_{-1.29}$	& $1.0^{+0.6}_{-0.6}$	& $83.62^{+0.14}_{-0.14}$ & $-$	& $48.7^{+56.0}_{-17.8}$& $0.732\pm0.014$ & $0.386\pm0.028$	& $19.09^{+64.80}_{-11.27}$	& EEE, FFF \\ [2pt]
	\footnotetext[1]{References for rotation periods: (a) \citet{2011ApJ...733..127S}; (b) \citet{2010ApJ...708..224H}; (c) \citet{2012MNRAS.422.1988A}; (d) \citet{2011AA...533A..88G}; (e) \citet{2011AA...529A..36S}; (f) \citet{2011AA...533A.130H}; (g) \citet{2011ApJS..197...14D}; (h) \citet{2013ApJ...775...54S}; (i) \citet{2012Natur.487..449S}}
	\footnotetext[2]{References for data: (A) \citet{2010AA...524A..25T}; (B) \citet{2013MNRAS.430.3422A}; (C) \citet{2010ApJ...708..224H}; (D) Cameron (\textit{priv comm.}); (E) \citet{2012MNRAS.422.1988A}; (F) \citet{2011AA...533A..88G}; (G) \citet{2013MNRAS.431..966T}; (H) \citet{2008AA...482L..21A}; (I) \citet{2011AA...533A.130H}; (J) \citet{2007MNRAS.379..773S}; (K) Wheatley (private communication); (L) \citet{2010AA...517L...1Q}; (M) \citet{2013AA...551A.108M}; (N) \citet{2011AA...528A..65M}; (O) \citet{2012ApJ...757..161T}; (P) \citet{2012MNRAS.423.1503B}; (Q) \citet{2011MNRAS.416.2593B}; (R) \citet{2010AA...519A..98B}; (S) \citet{2011AA...534A..16A}; (T) \citet{2014arXiv1402.1482A}; (U) \citet{2013AA...549A..18T}; (V) \citet{2012ApJ...760..139B}; (W) \citet{2011AJ....142...86E}; (X) \citet{2012AJ....143...81S}; (Y) \citet{2011AJ....141....8S}; (Z) \citet{2011AA...531A..40F}; (AA) \citet{2011PASP..123..547M}; (BB) \citet{2012MNRAS.426..739H}; (CC) \citet{2013AA...551A..73F}; (DD) \citet{2013AA...549A.134H}; (EE) \citet{2013AA...552A..82G}; (FF) \citet{2013AA...559A..36G}; (GG) \citet{2013AA...552A.120S}; (HH) \citet{2013arXiv1310.5630H}; (II) \citet{2012AA...537A.136G}; (JJ) \citet{2008ApJ...686..649J}; (KK) \citet{2013AA...551A..11M}; (LL) \citet{2010ApJ...720.1118B}; (MM) \citet{2011AA...533A.113M}; (NN) \citet{2011ApJ...742..116B}; (OO) \citet{2011ApJ...742...59H}; (PP) \citet{2012ApJ...757...18A}; (QQ) \citet{2010MNRAS.408.1689S}; (RR) \citet{2013ApJ...771...11A}; (SS) \citet{2012ApJ...759L..36H}; (TT) \citet{2010ApJ...718..575W}; (UU) \citet{2011AA...526A.130S}; (VV) \citet{2013MNRAS.430.2932M}; (WW) \citet{2012AA...544A..72L}; (XX) \citet{2013arXiv1310.5654A}; (YY) \citet{2011AJ....141...63W}; (ZZ) \citet{2009ApJ...696..241C}; (AAA) \citet{2009AA...503..601B}; (BBB) \citet{2009PASJ...61..991N}; (CCC) \citet{2010AA...516A..95H}; (DDD) \citet{2012ApJ...750..114F}; (EEE) \citet{2007ApJ...664.1190S}; (FFF) \citet{2008ApJ...682.1283W}; (GGG) \citet{2011AJ....141..179C}; (HHH) \citet{2013ApJ...775...54S} }
\end{longtable}
\endgroup
\end{landscape}
\twocolumn

\subsection{Isochrone ages}
\label{sec:isoimp}
The choice of isochrones being used can have a large impact on the derived properties of planetary systems. \citet{2009MNRAS.394..272S,2010MNRAS.408.1689S} suggests that multiple sets of isochrones should be used if at all possible, in preference to relying on a single formulation, as each model introduces its own systematic errors into the derived stellar parameters. I selected five sets of stellar models for my analysis: Padova isochrones \citep{2008AA...482..883M,2010ApJ...724.1030G}; YY isochrones (\citealt{2004ApJS..155..667D}); Teramo isochrones \citep{2004ApJ...612..168P}; Victoria-Regina isochrones (VRSS; \citealt{2006ApJS..162..375V}), and Dartmouth Stellar Evolution Database isochrones (DSED; \citealt{2008ApJS..178...89D}).

The main difficulty of isochrone fitting is that it is an attempt to fit a single point to a three-dimensional [$[M/H]$, $T_{\rm eff}$, $(\rho_{\rm s}/\rho_\odot)^{-1/3}$] parameter space in order to derive associated parameters (age and stellar mass). The problem can trivially be reduced to a two-dimensional one by considering only a single metallicity value at a time, which I achieve by neglecting the uncertainty in $[M/H]$. To convert between $[M/H]$ and $Z$, I use a value of $Z_{\odot}=0.0189$.

There are many possible fitting procedures. The simplest is to merely take the closest isochrone as the age of the system, but this often provides only crude estimates and has an accuracy that is constrained by the ages for which isochrones have been provided. A more involved approach would be to find the two closest isochrones and interpolate between them. Another alternative would be the Bayesian approach of \citet{2004MNRAS.351..487P}.

I have chosen to describe the [$T_{\rm eff}, (\rho_{\rm s}/\rho_\odot)^{-1/3}$] surface to which the stellar data is being fitted, and then to use this description to define a small plane over which I can interpolate the stellar data. For this purpose, I use a Delaunay triangulation, computed for a sub-region of the full isochrone parameter space that is centred on the measured stellar parameters. For details, please see Appendix\,\ref{sec:ages_triangulation}. Uncertainties in my interpolated ages are calculated by interpolating combinations of the $1\sigma$ limits on both $T_{\rm eff}$ and $(\rho_{\rm s}/\rho_\odot)^{-1/3}$ using the same procedure. 

\begin{figure}
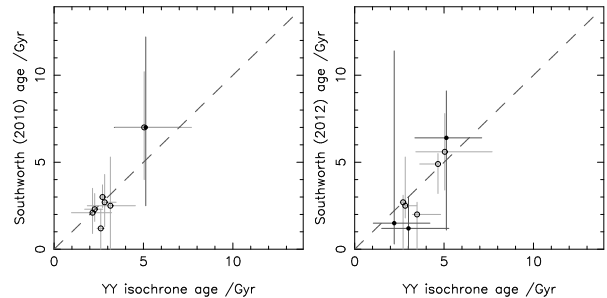

	\centering
	\begin{subfigure}{0.22\textwidth}
		\centering
		\includegraphics[width=\textwidth]{Figure2a}
	\end{subfigure}
	\begin{subfigure}{0.22\textwidth}
		\centering
		\includegraphics[width=\textwidth]{Figure2b}
	\end{subfigure}
	\caption{Left: ages from \citet{2010MNRAS.408.1689S} as a function of ages calculated using the YY isochrones in conjunction with my Delaunay triangulation interpolation technique. Right: ages from \citet{2012MNRAS.426.1291S} as a function of ages calculated using the YY isochrones in conjunction with my Delaunay triangulation interpolation technique. The dotted line denotes $y=x$. The maximum age on both axes is set to the age of the Universe. Direct measurements of the stellar rotation period were available for systems marked by solid symbols. Open symbols mark stars for which the rotation period has been derived using stellar parameters. In both cases, the ages are broadly similar, although the uncertainties that I find are significantly smaller. This arises due to \citeauthor{2012MNRAS.426.1291S}'s use of multiple sets of isochrones to determine systematic contributions to the uncertainties on his ages.}
	\label{fig:JKTCompare}
\end{figure}

\citet{2010MNRAS.408.1689S, 2012MNRAS.426.1291S} homogeneously studied large samples of exoplanet host stars, as part of which he carried out age determinations using a range of isochrones that included the YY isochrones. \citeauthor{2010MNRAS.408.1689S} uses a more traditional isochrone interpolation method, and as such these papers provide a reasonable comparison to my results. Cross-matching the results of those two papers to my own (see Figure\,\ref{fig:JKTCompare}) shows that there are eight systems in common in both cases. My results are generally compatible with those of \citet{2012MNRAS.426.1291S}, although it is immediately apparent that the uncertainties in my ages are smaller. I suspect that this partly results from \citeauthor{2010MNRAS.408.1689S}'s use of multiple isochrones to determine the systematic contribution to their age uncertainties, inflating their error bars somewhat. My uncertainties are also likely to be underestimated owing to my disregard for the uncertainty in metallicity. Comparing to \citet{2010MNRAS.408.1689S} I find similar ages for the younger stars, whilst for the two oldest systems in common (WASP-4 and WASP-5) I find younger ages, although the uncertainties on the ages are substantial.

\citet{2007ApJS..168..297T} studied a large sample of stars from the Spectroscopic Properties of Cool Stars (SPOCS) catalogue, calculating ages using the YY isochrones. Unfortunately, of the 1074 stars in their sample there are only 2 in common with this work -- HD\,209458 and HD\,80606. Our age estimates for HD\,209458 agree well, but for HD\,80606 the age that I calculate is much younger. As with the \citeauthor{2012MNRAS.426.1291S} studies, my uncertainties are smaller than those of \citeauthor{2007ApJS..168..297T}

\subsection{Gyrochronology calculations}
\label{sec:gyroimp}

\begin{figure*}
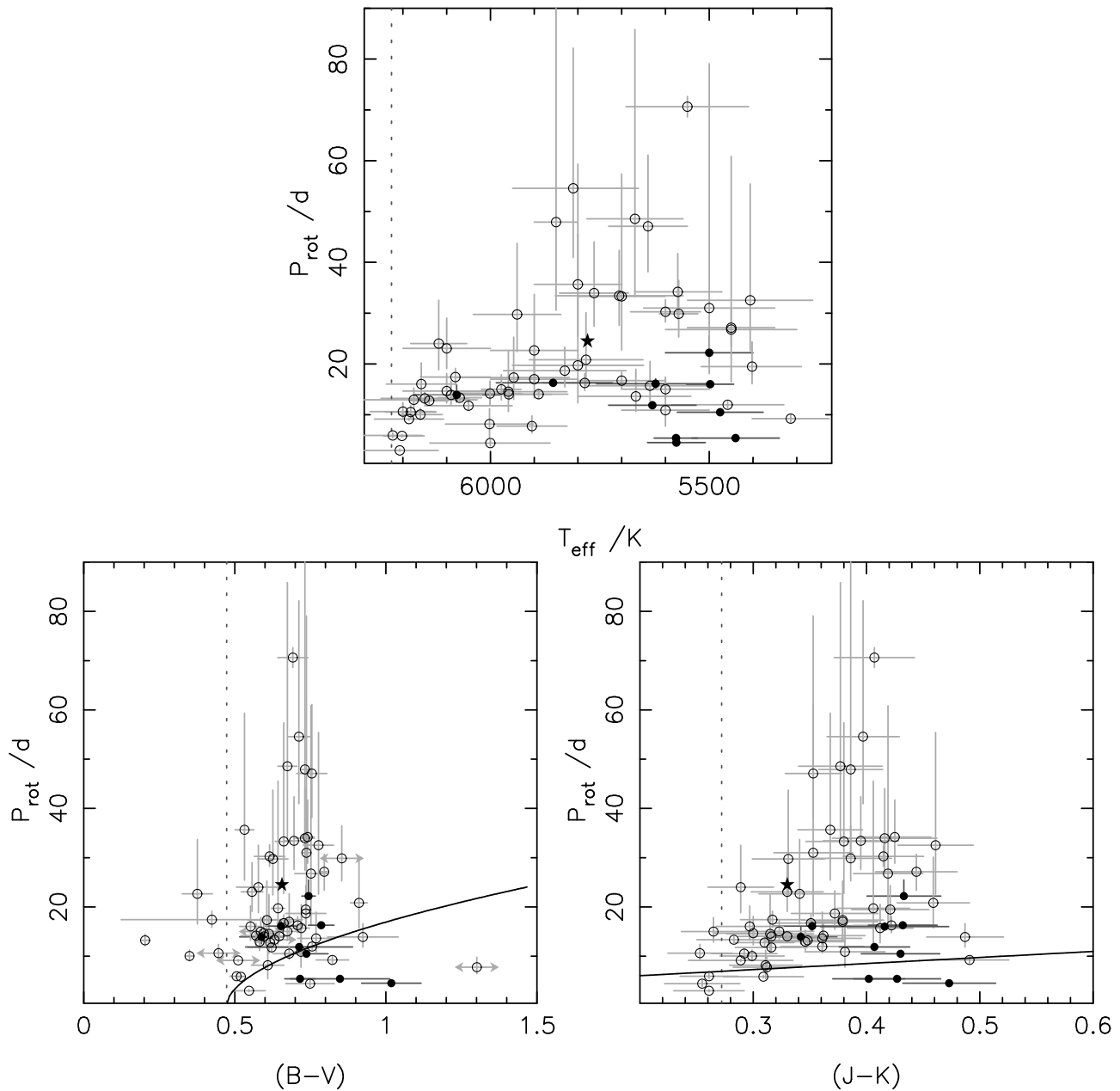

	\centering
	\begin{subfigure}{0.48\textwidth}
		\centering
		\includegraphics[width=\textwidth]{Figure3a}
	\end{subfigure}\\
	\begin{subfigure}{0.48\textwidth}
		\centering
		\includegraphics[width=\textwidth]{Figure3b}
	\end{subfigure}
	\begin{subfigure}{0.48\textwidth}
		\centering
		\includegraphics[width=\textwidth]{Figure3c}
	\end{subfigure}
	\caption{Upper: rotation period as a function of $T_{\rm eff}$. Lower left: Rotation period as a function of \textit{(B-V)} colour. The solid, black curve represents the period-colour-age relation from equation (\ref{eq:gyro1}), computed at the $0.625$\,Gyr age of the Hyades cluster. Lower right: Rotation period as a function of \textit{(J-K)} colour. The solid, black curve represents the period-colour relation from \citet{2009MNRAS.400..451C}, as in equation (\ref{eq:gyro2}). In all three figures, the position of the Sun is denoted by the solid star, and the location of the break in the Kraft curve is represented by the vertical dotted line. Solid symbols denote stars with measured rotation periods and open symbols mark stars for which the rotation period has been derived using stellar parameters.}
	\label{fig:ColPer}
\end{figure*}

I have used four different formulations of the $P_{\rm rot}$--colour--age relation to calculate ages for the systems in my sample. The first is from \citet{2007ApJ...669.1167B}, but uses updated coefficients from \citet{2009ApJ...695..679M} and \citet{2010AA...515A.100J} that were derived from studies of the M35 and M34 clusters, respectively:
\begin{equation}
\log \left(\frac{t}{\rm Gyr}\right)= \frac{\left[\log (P_{\rm rot}) - \log (0.770) - 0.553\log (B-V-0.472)\right]}{0.5344},
\label{eq:gyro1}
\end{equation} where $P_{\rm rot}$ is the stellar rotation period, and \textit{B} and \textit{V} are the stellar magnitudes in the Johnson B and V bands, respectively. The second formulation was derived from the period--colour relation for the Coma Berenices cluster by \citet{2009MNRAS.400..451C}:
\begin{equation}
t = 591\left[\frac{P_{\rm rot}}{9.30 + 10.39(J-K-0.504)}\right]^{1/0.56} {\rm Myr},
\label{eq:gyro2}
\end{equation}where $J$ and $K$ are the stellar magnitudes in the Johnson \textit{J} and \textit{K} bands respectively. The third formulation is similar to equation (\ref{eq:gyro2}), but was derived by \citet{2011MNRAS.413.2218D} using a study of the Hyades cluster:
\begin{equation}
t = 625\left[\frac{P_{\rm rot}}{10.603 + 12.314(J-K-0.570)}\right]^{1/0.56} {\rm Myr}.
\label{eq:gyro3}
\end{equation}The fourth and final formulation is that of \citet{2010ApJ...722..222B}:
\begin{equation}
t = \frac{\tau}{k_c}\ln\left(\frac{P}{P_0}\right) + \frac{k_I}{2\tau}\left(P^2-P_0^2\right) {\rm Myr},
\label{eq:gyro4}
\end{equation} where $k_c=0.646$\,d\,Myr$^{-1}$ and $k_I=452$\,Myr\,d$^{-1}$ \citep{2010ApJ...721..675B}, $P_0$ is the rotation period of the star at time $t=0$ (assumed to be $1.1$\,d, the initial period of the calibrated solar-mass model in \citealt{2010ApJ...722..222B})\footnote{Note that the choice of $P_0$ can affect the gyrochronology age obtained, particularly for younger stars.}, and $\tau$ is the convective turnover time-scale. For each system, I determine $\tau$ using table\,1 of \citet{2010ApJ...721..675B} and the star's effective temperature.

Broad-band colour indices for the WASP systems were derived using magnitude data from the AAVSO Photometric All-Sky Survey (APASS; \citealt{2012JAVSO..40..430H}), accessed through the UCAC4 catalogue \citep{2013AJ....145...44Z}, and 2MASS \citep{2006AJ....131.1163S} for the \textit{(B-V)} and \textit{(J-K)} colours, respectively. For systems with available stellar rotation period measurements, I created a Gaussian distribution with mean and variance set to the known period and $1\sigma$ error, respectively. The distribution was sampled $10^4$ times, and for each sampling I calculated age estimates using equations (\ref{eq:gyro1})--(\ref{eq:gyro4}). Final ages for each method were taken to be the median of the appropriate set of results, with $1\sigma$ uncertainties set to the values which encompassed the central $68.3$\,percent of the data set.

For the majority of the planetary systems in my sample there exists no direct measurement of the stellar rotation period. I therefore sampled Gaussian distributions for the projected stellar rotation $v\sin I_{\rm s}$, the orbital inclination $i_{\rm orb}$, and the stellar radius, $R_{\rm s}$, $10^4$ times as described above. For each set of sampled data I calculated the rotation period using
\begin{equation}
P_{\rm rot} = \frac{2{\rm \pi}R_s}{v\sin I_{\rm s}}\sin i_{\rm orb}.
\label{eq:Prot}
\end{equation}I assumed that the systems are aligned along the line of sight such that the inclination of the stellar rotation axis to the line of sight $I_{\rm s}=i_{\rm orb}$, and used the gyrochronology equations to calculate age estimates. Final age estimates were calculated as above.

Nine of the systems in my sample have directly measured rotation periods available. For these systems, I also calculated the rotation period, allowing me to compare the derived periods to the measured values. In five out of the nine cases, the two periods agree, although in the case of WASP-46 this is due to the substantial uncertainty in the derived period. For the remaining three systems, the periods disagree by more than $5\sigma$. This is likely due to misalignment of the stellar rotation axis along the line of sight such that $v\sin I_{\rm s}$ is not a good representation of the true rotation speed of the stars involved. Three of the systems for which the two periods disagree have a derived period that is longer than the measured period, supporting the case for a misaligned star. The exception is WASP-19, for which the derived period is significantly shorter than the measured period for reasons unknown.

It is also possible that differential rotation in the star has led to the rotation period being measured at a latitude other than the stellar equator. Evidence for differential rotation has in fact been observed for CoRoT-2 \citep{2009AA...506..263F,2010AA...514A..39H}, with possible indications also present for WASP-19 \citep{2011ApJ...730L..31H,2013MNRAS.428.3671T}, but the scale of the effect is insufficient to explain the discrepancy between the measured and derived rotation periods. Misalignment along the line of sight thus seems a more probable explanation for discrepant derived periods among my stellar sample, but without measurements of the stellar inclination the reliability of the derived periods is difficult to ascertain. Assuming $i_{\rm orb}=I_{\rm s}$ initially seems reasonable, but it has been shown that some systems will have stellar inclinations such that this assumption is invalid \citep{2010ApJ...719..602S}. 

In Figure\,\ref{fig:ColPer}, I plot effective temperature, \textit{(B-V)} colour, and \textit{(J-K)} colour as functions of rotation period, and overplot both the position of the break in the Kraft rotation period curve \citep{1967ApJ...150..551K} and relevant relationships between colour and period. As already noted, my sample is selected on $T_{\rm eff}$ with an upper limit that approximates to the temperature at which the Kraft break occurs. This is clear from the upper panel of Figure\,\ref{fig:ColPer}, with only nine systems having uncertainties such that they might lie slightly `above' the break. The location of the break in \textit{(J-K)} space is also close to the edge of the sample; five systems appear to have \textit{(J-K)} colour such that they lie `above' the break, though the uncertainties are such that that number could be anywhere between $0$ and $11$ . But when translated into \textit{(B-V)} colour space, the Kraft break seems to shift within the sample, with three systems displaying lower \textit{(B-V)} colour index than the position of the break even when uncertainties are accounted for. However, these three systems are the same as those that were out of place in Figure\,\ref{fig:CMD}; checking other sources suggests that the APASS calibration for these three stars is likely to be inaccurate, and I therefore exclude them when analysing ages calculated using equation (\ref{eq:gyro1}).

\section{Comparing the age calculation methods}
\label{sec:results}
Although I have carried out similar analyses for all $20$ combinations of the isochrones and gyrochronology relations mentioned above, in the discussion that follows I will concentrate on the results obtained using the YY isochrones and the \citet{2010ApJ...722..222B} gyrochronology formulation described by equation (\ref{eq:gyro4}). The comparison utilizes only those systems with valid results for both methods. The maximum permitted age for any star was set to the current best estimate of the age of the Universe \citep[and other papers in the series]{2013arXiv1303.5062P}, and systems with calculated ages greater than this were disregarded. This is perhaps a somewhat unrealistic upper bound; the age of the Galactic disc might be more suitable (and is thought to be somewhat younger than the Universe), but introduces its own set of problems. Do the thick and thin discs have the same age, and if not, which should be used? Or should the sample be split up by population, and if so how would that be done (disc component membership is a difficult attribute to characterize)? For simplicity, I have stuck to the age of the Universe.

The null hypothesis of this work is that the two age calculation methods are equally accurate, and therefore that the ages calculated using the two different methods will agree. This may not always be true on a case-by-case basis, but when viewed as an overall sample, then agreement is the expected outcome.

\subsection{Isochrones versus gyrochronology}
\label{sec:agescatter}
As a starting point, I plot gyrochronology age as a function of isochrone age. If the two methods provided similar answers, I would expect a tightly correlated sequence centred on the line ${\rm age}_{\rm gyro}={\rm age}_{\rm iso}$ (within errors). However, there appears to be a preponderance of points lying towards the isochronal side of the line, suggesting that isochrone fitting tends to return ages that are older than those preferred by gyrochronological methods. Neglecting uncertainties, there are twice as many systems for which the isochrone age is older than for which the gyrochronology age is older; this ratio increases once uncertainties are taken into account, in a large part owing to the large uncertainties on my calculated gyrochronology ages. But the number of systems with error ellipses consistent with equal ages is, at $30$ systems, more than half of the $57$ systems for which valid ages were returned by both methods. This is a significant fraction of the sample, and indeed one would hope that this would be the case given the stated null hypothesis. However, the number of systems for which the isochrone age is still greater than the gyrochronology age once the uncertainties are taken into account is $22$, whilst the converse case includes only $5$ systems. 

\begin{figure}
	\centering
 	\includegraphics[width=0.48\textwidth]{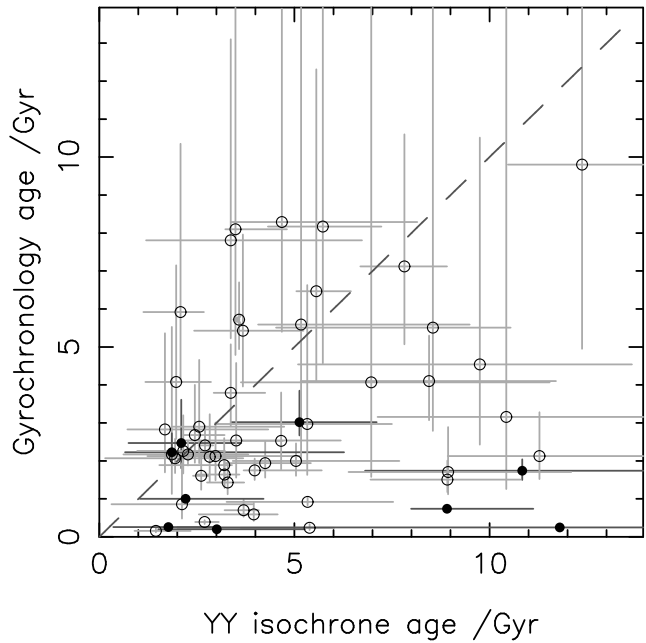}
	\caption{Gyrochronology age, calculated using equation (\ref{eq:gyro4}), as a function of isochrone fitting age, found using the YY isochrones. The dashed line denotes $y=x$; systems clustered around this line show similar age values for different methods of calculation. The maximum age on both axes is set to the age of the Universe. Direct measurements of the stellar rotation period were available for systems marked by solid circles. For systems marked by open circles, $P_{\rm rot}$ was derived from $v\sin I_{\rm s}$ and $R_{\rm s}$ according to equation (\ref{eq:Prot}). It appears that the gyrochronology ages have a slight tendency to be younger than isochrone-fitting ages, particularly for systems with measured rotation periods.}
	\label{fig:G4vsY}
\end{figure}

Another interesting facet of Figure\,\ref{fig:G4vsY} is the distribution of the points along both axes. Just over half of the systems lie within a region defined by ${\rm age}_{\rm gyro}<4$\,Gyr and ${\rm age}_{\rm iso}<6$\,Gyr. This is not entirely surprising given the region of parameter space to which I have restricted the study. Rough estimates of $\tau_{\rm MS}$ for stars at the limit of my parameter space are $\tau_{\rm MS}=3.5$\,Gyr for an F7 star and $\tau_{\rm MS}=11.4$\,Gyr for a G9 star (using masses from table B1 of \citealt{2008oasp.book.....G}). A drop-off after roughly 4 Gyr is consistent with this, as systems at the hotter end of the parameter range start to evolve off the MS, and are therefore no longer targeted by transit search programs. Including uncertainties in this analysis lowers the number of systems that are definitively within this high-density region to $21$, with a further $19$ which have error ellipses at least partially within this region of parameter space.

In terms of the different methods, $60$\,percent of the gyrochronology estimates are less than $4$\,Gyr, with possible stellar ages ranging from $0.3$\,Gyr up to the age of the Universe. For the isochrone-fitting estimates, $54$\,percent are younger than $6$\,Gyr, with the estimates covering a similar range. It therefore seems, at first glance, that gyrochronology tends to return stellar age estimates which are slightly biased towards younger ages than the results from isochrone fitting. Whilst this conclusion is tempered somewhat by the magnitude of the uncertainties on the ages that I have calculated, particularly for the gyrochronology ages, it may be true even accounting for these.

A 2D Kolmogorov--Smirnov (KS) test on the two data sets indicates that there is a less than $1$\,percent probability of the two having a common parent distribution, but this fails to account for the uncertainties in my ages. I therefore evaluate the $\chi^2$ goodness of fit of my data to the line ${\rm age}_{\rm Gyro} = {\rm age}_{\rm Iso}$,
\begin{equation}
	\chi^2 = \Sigma\frac{({\rm age}_{\rm Gyro} - {\rm age}_{\rm Iso})^2}{\sigma_{\rm Gyro}^2 + \sigma_{\rm Iso}^2},
	\label{eq:chi2dof}
\end{equation} where $\sigma_{\rm Gyro}$ and $\sigma_{\rm Iso}$ are the uncertainties in each value of the gyrochronological and isochrone-fitting ages, respectively. I find $\chi^2=273.4$, with a reduced value of $\chi^2_{\rm red}=4.1$, suggesting that my ages are a poor match for the null hypothesis. The \textit{P}-value for this result is $P(\chi^2)\sim0$, a strong indication of significance.

\begin{figure}
	\centering
	\includegraphics[width=0.48\textwidth]{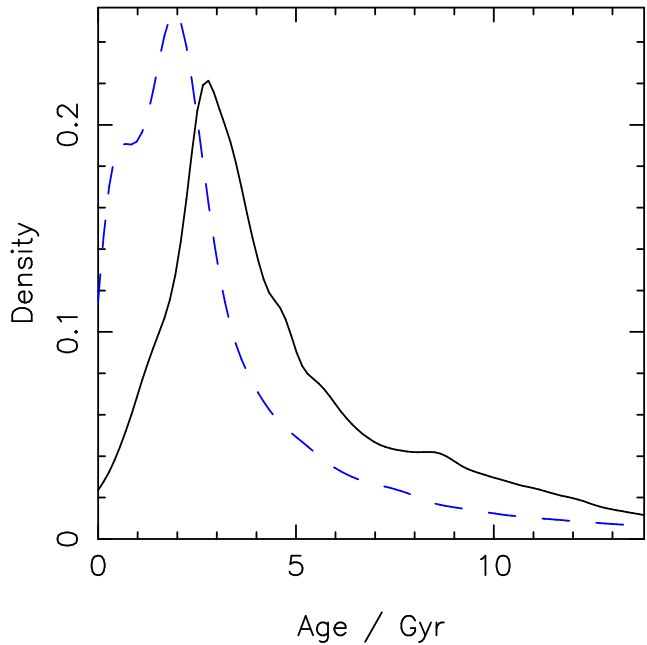}
	\caption{KDEs for the results that I obtained from isochrone fitting to the YY isochrones (solid, black line), and from gyrochronology using equation (\ref{eq:gyro4}) (dashed, blue line). The highest peaks of the two distributions are at $2.8$\,Gyr for the isochrone-fitting results and $1.5$\,Gyr for the gyrochronology results. This suggests a small overarching offset between the sets of age estimates provided by the two methods.}
	\label{fig:G4vsY_dist}
\end{figure}

To further examine the different distributions I computed kernel density estimates (KDEs; \citealt{parzen1962,rosenblatt1956}) for the two data sets, additionally disregarding systems for which one or both of the two methods returned only an upper or lower limit on the age. One of the advantages of this visualization method compared to cumulative probability distributions or histograms is that it intrinsically accounts for the uncertainties in the measured parameters, giving a more accurate idea of the shapes of the distributions and allowing more concrete comparison between them. For each system, I took $10^4$ random samples from a normal distribution with a mean of $0$ and a standard deviation of $1$,  scaling these random numbers according to the system's age and $1\sigma$ uncertainties. Combining these sets of sampled ages across all of the systems being examined, I used Scott's Rule \citep{1992mde..book.....S} to compute the KDEs. These can be  seen in Figure\,\ref{fig:G4vsY_dist}, sampled at $100$ points evenly spaced between $0.0$\,Gyr and the age of the Universe.

The highest peaks of the two KDEs lie at $2.8$\,Gyr and $2.0$\,Gyr for isochrone fitting and gyrochronology, respectively, again suggesting that there may be a slight difference in the age estimates being returned by the two methods. This visualization technique also provides another look at the different regions of parameter space occupied by the two sets of results, as the relative heights and widths of the two peaks again indicate that the gyrochronology results are concentrated in a slightly smaller region of parameter space than the isochrone-fitting results.

\subsection{$\Delta{\rm age}$ analysis}
\label{sec:delageprob}
\begin{figure}
	\centering
	\includegraphics[width=0.48\textwidth]{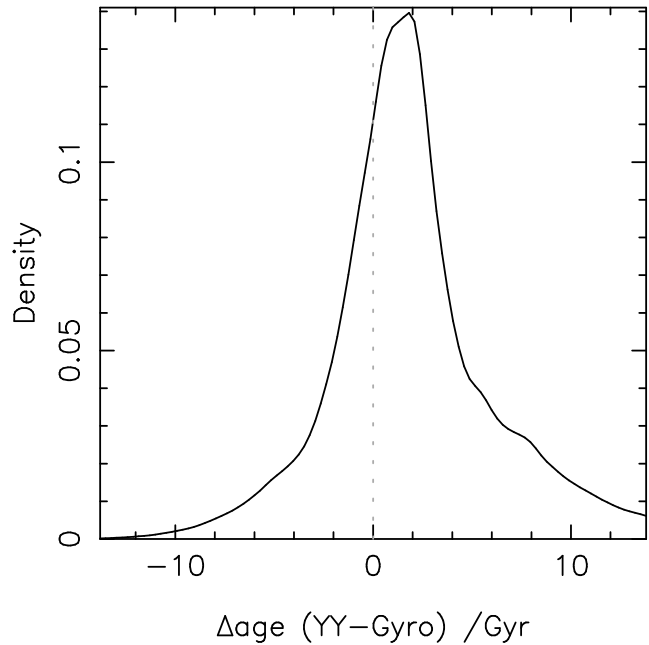}
	\caption{KDE for the difference between the age results obtained by isochrone fitting using the YY isochrones, and by gyrochronology using equation (\ref{eq:gyro4}). The vertical dotted line denotes $\Delta{\rm age}=0$, and the peak of the KDE is offset towards positive $\Delta{\rm age}$. This again suggests that isochrone fitting is returning ages which are slightly older than those from gyrochronology. }
	\label{fig:delG4Y_cumdist}
\end{figure}

To further investigate this bias, I calculated $\Delta{\rm age}={\rm age}_{\rm Iso}-{\rm age}_{\rm Gyro}$ for each of the systems in my sample and computed a KDE for the set of results. Figure\,\ref{fig:delG4Y_cumdist} shows a small apparent offset towards positive $\Delta{\rm age}$, in line with the suggestion from the previous section that isochrone fitting is returning ages which are slightly older than those from gyrochronology. The peak of the KDE lies at $1.8$\,Gyr, and the average upper and lower error bars on $\Delta{\rm age}$ are $4.0$ and $2.1$\,Gyr, respectively, so this is an inconclusive $0.9\sigma$ effect. However, the KDE is asymmetrical, with a narrower peak but broader shoulder in positive $\Delta{\rm age}$ than in negative $\Delta{\rm age}$; comparison to Figure\,\ref{fig:G4vsY_dist} shows that this derives from the isochrone-fitting KDE. This matches the distribution of the data in Figure\,\ref{fig:G4vsY}: there are more data for which the isochronal age is older than the gyrochronological age, but the uncertainties are large enough that they dilute the effect.

Although the effect in $\Delta{\rm age}$ is small, the comparison of the individual methods suggests that there might be a disagreement between the ages that are produced by gyrochronology and isochrone fitting. Does this possible discrepancy correlate with a physical parameter in the systems that I am studying? Is it that isochrone fitting is overestimating ages, or that gyrochronology is underestimating ages (or a combination of the two)? Figure\,\ref{fig:G4vsY} certainly seems to imply the former, as the systems with measured rotation periods, and thus the most reliable gyrochronology ages, all show a tendency towards an older isochronal age, in some cases with strong significance. This could be an indication that my new method for determining isochrone ages is overestimating the ages of my systems; with the comparison sample that is available (see Figure\,\ref{fig:JKTCompare}), it seems that this is not the case on average, although the comparison is very limited in scope. As noted in Section\,\ref{sec:gyro} though, gyrochronology is only applicable if no external factors act to modify the natural stellar spin-down. If hot Jupiter host stars are rotating more rapidly than expected, then their age would be underestimated.

\subsection{The influence of tidal interactions}
\label{sec:age_tides}
One possibility might be that the spin rate of the star is being modified somehow, and angular momentum exchange between the star and the planet's orbit provides one route by which such a scenario might occur. The chief method of angular momentum exchange within planetary systems is through tidal interaction, which has well-documented consequences for stellar spin. For this work I am interested in the possibility of a link between the strength of the tidal interactions and the magnitude of the difference between my age estimates.

To investigate this I calculated the theoretical tidal time-scale for each of my systems using
\begin{equation}
\frac{1}{\tau_{\rm CE}} = \frac{1}{10\times10^9} q^2 \left(\frac{a/R_{\rm s}}{40}\right)^{-6} {\rm yr},
\label{eq:ttauCE}
\end{equation}where $q=M_{\rm p}/M_{\rm s}$ is the ratio of the planetary and stellar masses, $a$ is the planet's orbital semi-major axis, and $R_{\rm s}$ is the stellar radius \citep{2012ApJ...757...18A}. $\tau_{\rm CE}$ is the tidal time-scale for alignment through dissipation in convective envelopes; since I apply an upper limit for my sample at the Kraft break temperature of $T_{\rm eff}=6226$\,K, I neglect the time-scale for tidal dissipation in radiative stars.

\begin{figure}
	\centering
	\includegraphics[width=0.48\textwidth]{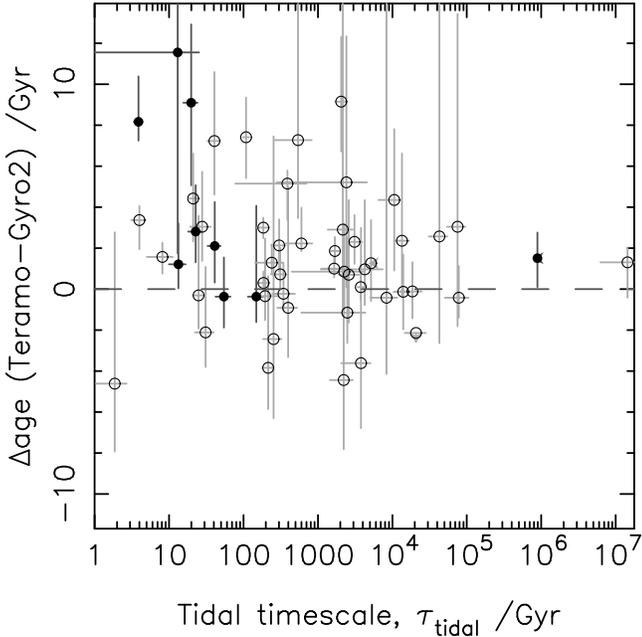}
	\caption{$\Delta{\rm age}$ as a function of $\tau_{\rm tidal}$, the tidal realignment time-scale. The shorter the time-scale, the stronger the tidal interactions within the system, and the greater the angular momentum exchange. No trend is apparent between $\Delta{\rm age}$ and $\tau_{\rm tidal}$, although two of the systems with measured rotation period (solid symbols) show significantly greater age difference and significantly shorter tidal time-scale than the other systems for which $P_{\rm rot}$ has been measured. The horizontal dashed line marks $\Delta{\rm age}=0$. }
	\label{fig:delG4Y-ttide}
\end{figure}

In Figure\,\ref{fig:delG4Y-ttide}, I plot $\Delta{\rm age}$ as a function of tidal time-scale. If angular momentum exchange is the cause of the discrepancy between the two age estimation methods, then I would expect the difference to be greatest for systems with the shortest tidal time-scales (i.e. the strongest tides). Unfortunately the evidence is inconclusive owing to the size of the uncertainties on $\Delta{\rm age}$, which means that any conclusion would be tentative at best. Any evidence for a trend is also countered by the number of systems for which $\Delta{\rm age}$ is apparently negative, although it is worth noting that several of these systems have substantial uncertainties such that they are consistent with $\Delta{\rm age}=0$.

Considering only the stars in my sample with directly measured rotation period (the solid data in Figure\,\ref{fig:delG4Y-ttide}) reveals a possible trend, with two systems exhibiting an age difference of a few Gyr, and tidal time-scales close to the minimum value for the sample. A sample size of only nine systems means that this is far from clear however, and any possible trend hinges on two of the systems. It is interesting, however, to further consider the WASP-19 system, which is the system with the shortest tidal time-scale, $3.89$\,Gyr, of those for which the rotation period has been measured.

WASP-19\,b has the shortest orbital period of the WASP planets, and has an orbital semi-major axis of only $0.01653\pm0.00013$\,au. $\Delta{\rm age}=8.17^{+2.21}_{-0.92}$\,Gyr for this system, implying that the gyrochronology age is being underestimated. \citet{2011MNRAS.415..605B} investigated the possibility of tidal interactions, finding that it is possible that the star is undergoing tidal spin-up which would explain the underestimation of the gyrochronology age (assuming that the isochrone age is correct).

\subsubsection{Tidal effects in the WASP-19 system}
\label{sec:wasp19}
To check the plausibility of tidal spin-up I consider a range of evolutionary scenarios for the WASP-19 system. A full investigation of the tidal effects is beyond the scope of this paper, but it is simple to compare the rotation periods produced by different combinations of parameters at the expected age of the system.

I used the tidal equations and integration procedure described in \citet{2011MNRAS.415..605B}, starting from a set of defined initial conditions. I fixed the stellar and secondary body tidal quality factors at the values determined for the WASP-19 system by \citeauthor{2011MNRAS.415..605B}, set the orbital eccentricity to $0$, and set $P_{0}=1.1$\,d to determine the initial stellar rotation frequency. The secondary body's spin was assumed to be synchronized to it's orbit. $a_0$ and $M_2$ were varied to produce different rotational histories for the star, with $R_2 = R_{\rm Jup}$ for planetary mass companions, and $R_2=R_{\odot}$ for stellar mass companions.

I first calibrated the simulations by turning off tidal interactions such that the only influence on the rotation of the primary body was magnetic braking. At the age of $8.91^{+2.21}_{-0.92}$\,Gyr provided by the YY isochrones, this gives a rotation period of $33.1^{+5.0}_{-3.0}$\,d and a gyrochronology age of $3.4^{+0.4}_{-0.5}$\,Gyr using equation (\ref{eq:gyro4}). There is still a substantial offset between this and the isochronal age estimate, with isochrone fitting again overestimating the age; indeed the quoted isochronal age lies towards the upper bound of that given in \citet{2011MNRAS.415..605B}. The rotation period is also significantly longer than the measured period of $10.2\pm0.5$\,d., which is at odds with the period derived using equation (\ref{eq:Prot}) in Section\,\ref{sec:gyroimp}. This might suggest that the initial rotation rate of the star is poorly estimated, but I found that changing the initial rotation period had little effect on the rotation period derived using this calibration scenario. This is unsurprising given previous gyrochronology work which shows that stars tend to converge to a single period--colour--age relation within a few hundred Myr.

For an initial separation of only $0.05$\,au, none of the planetary mass secondaries survived to the isochronal age for the system listed in Table\,\ref{tab:data}; all migrated inwards to the Roche limit before this, causing significant spin-up of the host as they did so such that the stellar rotation period at time of destruction was consistent with the measured value. Increasing the initial separation to $a_0=0.0625$\,au revealed that secondaries of mass $0.5$\,$M_{\rm Jup}\leq M_2\leq1.5$\,$M_{\rm Jup}$ produced no difference in gyrochronology age compared to the calibration case, although the larger masses did produce marginally significant differences in rotation period of $1-2$\,d. Increasing the initial separation still further to $0.075$ and $0.10$\,au showed that masses of $M_2\geq5$\,$M_{\rm Jup}$ and $M_2\geq15$\,$M_{\rm Jup}$ were required to produce the same, minimally measurable differences in rotation period and age. For comparison, the measured parameters for WASP-19\,b are $M_{\rm p}=1.14\pm0.07$\,$M_{\rm Jup}$ and $a=0.0164^{+0.005}_{-0.006}$\,au.

These results imply that the initial separation must have been $<0.0625$\,au if both the isochronal age and measured rotation period are correct, as the only way to reconcile the two is through stellar spin-up. Note also that the currently observed semi--major axis of the system is very difficult to replicate in this simplistic model, with only models causing spin-up being able to match the current orbit.

Using a stellar mass secondary showed that for $a_0<0.25$\,au the rotation of the primary at the isochronal age was substantially faster than measured, with commensurately younger gyrochronology age estimates; the difference between the simulated and measured rotation period \textit{decreased} as the initial separation increased. In any case, simply replacing the planet with a secondary body of stellar dimensions would have a much more severe effect on the rotation, and thus the derived age, of WASP-19\,A.

\subsection{A link with spectral type?}
\label{sec:age_sptype}
\begin{figure}
	\centering
	\includegraphics[width=0.48\textwidth]{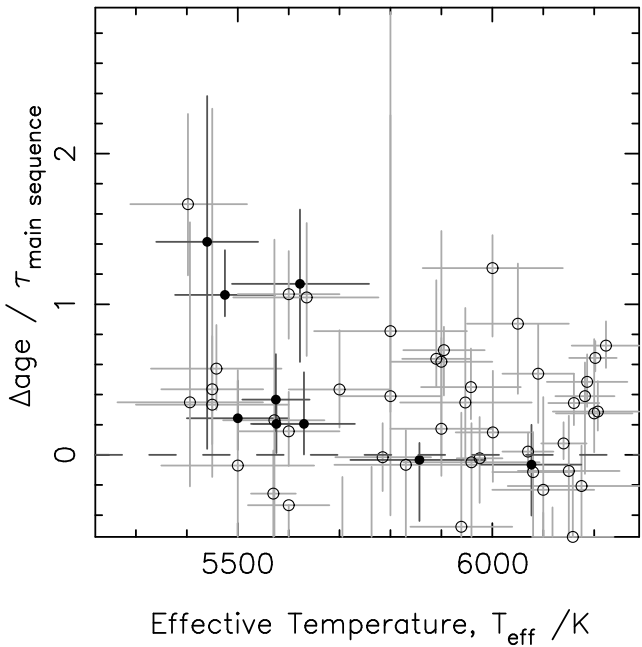}
	\caption{$\Delta{\rm age}$/MS lifetime as a function of $T_{\rm eff}$, the stellar effective temperature. No clear trends are present. Consideration of only those stars with measured rotation period (solid symbols) shows a possible slight trend for $\Delta{\rm age}$ to increase with decreasing $T_{\rm eff}$, towards later spectral types.}
	\label{fig:delG4Y-teff}
\end{figure}

Having investigated the possibility that the gyrochronology results are too young, I turn my attention to the alternative possibility that some of the isochrone fitting results are too old. This might manifest as a bias with spectral type. To see whether any such trend is exhibited in my sample, I divide $\Delta{\rm age}$ by the MS lifetime of the stars, and plot the resulting age ratio as a function of effective temperature in Figure\,\ref{fig:delG4Y-teff}. The sample as a whole shows little in the way of a trend, again due to the magnitude of the uncertainties in the age ratios. However, if I consider only the stars with measured rotation periods (the solid data), then there might be a small trend for the age ratio to increase as $T_{\rm eff}$ decreases. This conclusion is driven entirely by two of the stars in the already small set however, and should only be considered as a possibility until further rotation periods are obtained and used to recalculate gyrochronology ages for additional stars in my sample.

\begin{figure}
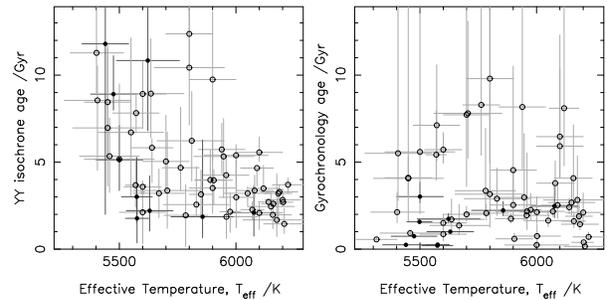

	\centering
	\begin{subfigure}{0.22\textwidth}
		\centering
		\includegraphics[width=\textwidth]{Figure9a}
	\end{subfigure}
	\begin{subfigure}{0.22\textwidth}
		\centering
		\includegraphics[width=\textwidth]{Figure9b}
	\end{subfigure}
	\caption{Age as a function of stellar effective temperature. Solid symbols mark systems with measured rotation periods. Left: ages calculated through isochrone fitting using the YY isochrones. Right: ages calculated through gyrochronology using equation (\ref{eq:gyro4}). Whilst the ages from isochrone fitting show similar trends with $T_{\rm eff}$ as $\Delta{\rm age}$, the ages from gyrochronology show no such trends. }
	\label{fig:G4vsY-teff}
\end{figure}

From Figure\,\ref{fig:G4vsY-teff}, it seems that there might be some differences in the dependence on $T_{\rm eff}$ between the two methods. The gyrochronology results are distributed evenly across the temperature range that I am considering, although the uncertainty on the age estimates increases dramatically for ${\rm age}\gtrsim4$\,Gyr. In contrast, the isochrone results seem to show a trend with $T_{\rm eff}$, with the oldest stars also being the coolest; this is a selection effect, as old, hot stars will have evolved off the MS, and the majority of my sample consists of hot Jupiters discovered by transit surveys which, as I have already remarked, select against older, evolved stars. There is also a noticeable trend in the uncertainties on the isochrone results, with the younger, hotter stars exhibiting more precise ages. This concurs with a study by \citet{2004MNRAS.351..487P}, who noted that the size of the observational uncertainties relative to the separation of the isochrones was an important parameter for isochrone fitting, and one which was most favourable for young, hot, early-type systems. The trend in $\Delta{\rm age}$ with temperature, if it exists, therefore seems to result from the isochronal results.

\subsection{Planet hosts in stellar clusters}
\label{sec:clusters}
The nature of isochrone dating itself could also be a factor. The method relies on the choice of isochrones, which in turn relies on having an accurate distance estimate to the star in question. With field stars such as those that comprise my sample, this is inherently very difficult, and distance data for the stars that I am studying are sparse -- only four have \textit{Hipparcos} parallax measurements. Stars in clusters provide more suitable targets, as they will usually have a well-defined distance and an accepted age. Unfortunately, the number of cluster stars that are known to host planets is small, and the number for which all of the data required for my age estimation techniques are available is smaller still.

I was able to calculate ages for five planet-hosting cluster stars: HD\,285507 in the Hyades \citep{2013arXiv1310.7328Q}; two stars in Praesepe \citep{2012ApJ...756L..33Q}, and two stars in NGC\,6811 \citep{2013Natur.499...55M}. In each case there was no discussion of stellar rotation in the context of the stellar cluster, which I take to mean that the rotation of the stars concerned is typical. Table\,\ref{tab:cluster} displays the age estimates that result.

HD\,285507 is cool when compared to my sample, and the isochronal method duly struggled, finding an age of $7.5^{+7.9}_{-7.4}$\,Gyr. Equation (\ref{eq:gyro4}) also overestimates the age of the star, albeit by a much narrower margin, finding $0.70\pm0.01$\,Gyr compared to the cluster age of $0.625\pm0.050$\,Gyr. This is particularly interesting, as the rotation period of the star has been measured through characterization of its photometric variability \citep{2011MNRAS.413.2218D}, so the gyrochronology relation should give good agreement with the age of the cluster. \citep{2013arXiv1310.7328Q} suggest that the orbit of the planet might have been circularized, indicating past tidal interaction which might have affected the star's rotation.

Both of the Praesepe stars fall within the bounds of my sample's parameter space. For both stars, I found good agreement with the cluster age of $0.578\pm0.049$\,Gyr using equation (\ref{eq:gyro4}), whilst isochrone fitting only returned upper limits on the age of the stars concerned. \citet{2011MNRAS.413.2218D} provide a plot of rotation period as a function of \textit{(J-K)} colour. Comparing my derived rotation periods and \textit{(J-K)} values to this plot shows that Pr0211 lies nicely on the sequence that they find, and using equation (\ref{eq:gyro3}) I find an age of $0.55^{+0.15}_{-0.13}$\,Gyr for the system, in agreement with the cluster age as expected. Pr0201 lies to the left of their data, but extrapolating the plot implies that it too is in rough agreement with their period-colour sequence. I derive an age of $0.41^{+0.10}_{-0.09}$ using equation (\ref{eq:gyro3}), very slightly underestimated compared to the cluster age. Equation (\ref{eq:gyro2}), which also uses \textit{(J-K)}, gives similar results for both stars.

The planet-hosting stars in NGC\,6811 also both fall within my parameter space, but equation (\ref{eq:gyro4}) only gives agreement with the cluster age in the case of Kepler-66; for Kepler-67, gyrochronology gives a younger age than expected. As with the Praesepe stars, I obtain only upper limits for the stellar ages using isochrone fitting with the YY models.

The case of HD\,285507 in particular highlights the challenges involved in isochrone fitting. Either the wrong isochrone has been selected (not impossible, even with a known distance, as extinction must also be taken into account), or the method is struggling to deal with the young age owing to the close packing of the isochrones at the age of the cluster. The overestimation of the age using gyrochronology is intriguing, and might point towards an overactive star that is losing angular momentum more quickly than expected.

\begin{table*}
	\caption{Age estimates obtained for the five planet-hosting cluster stars discussed in Section\,\ref{sec:clusters}.}
	\label{tab:cluster}
	\begin{tabular}{llllll}
		\hline \\
		Cluster		& Cluster age (Gyr)	& Star			& $T_{\rm eff}$ (K)		& YY age (Gyr)			& ${\rm Gyro}$ ${\rm age}_4$ (Gyr)	\\ [2pt]
		\hline \\
		Hyades		& $0.625\pm0.025$ & HD\,285507		& $4503^{+85}_{-61}$	& $7.5^{+7.9}_{-7.4}$	& $0.70\pm0.01$				\\ [2pt]
		Praesepe		& $0.589\pm0.049$	& Pr0201			& $6174\pm50$		& $<2.44$				& $0.63^{+0.20}_{-0.15}$			\\ [2pt]
					& 				& Pr0211			& $5326\pm50$		& $<5.54$				& $0.55^{+0.14}_{-0.11}$			\\ [2pt]
		NGC\,6811	& $1.00\pm0.17$	& Kepler-66		& $5962\pm79$		& $<3.35$				& $1.01^{+0.15}_{-0.11}$			\\ [2pt]
					&				& Kepler-67		& $5331\pm63$		& $<5.87$				& $0.70\pm0.02$				\\ [2pt] 
		\hline \\
	\end{tabular}
\end{table*}

\section{Systems with measured spin--orbit angles}
\label{sec:age_lambda}
An area of planet research where tides are widely thought to play a role is the angle of alignment, $\lambda$, between the stellar spin axis and the planet's orbital axis. Examining a sample of planetary systems for which $\lambda$ has been measured might therefore be able to shed more light on whether tidal interactions influence gyrochronology age estimates for planet-hosting stars.

The subject of spin-orbit alignment is comprehensively covered elsewhere, so I will not dwell on it here. Suffice it to say that for planetary systems we have measured a variety of angles between the rotation axis of the host star and the orbital axis of the planet. Once the angle has been measured, the system is classified as `aligned' or `misaligned' according to some criterion. Here, I will be using that of \citet{2010ApJ...718L.145W}, who define a system as `misaligned' if $\lambda\geq10^\circ$ to $>3\sigma$. It is thought that tidal interactions are involved in determining whether a system is `aligned' or `misaligned', with tidal realignment of the stellar spin axis to the planet's orbital axis thought to produce the evolution of orbits from one group to the other.

In this section, I repeat my previous analysis, this time considering only those systems for which $\lambda$ has been measured. These were selected using the Holt--Rossiter--McLaughlin data base of Ren\'{e} Heller\footnote{www.physics.mcmaster.ca/$\sim$rheller/}, as of 2013 October 24. The reduced sample consists of $31$ systems, $8$ of which are classified as misaligned.

\begin{figure}
	\centering
 	\includegraphics[width=0.48\textwidth]{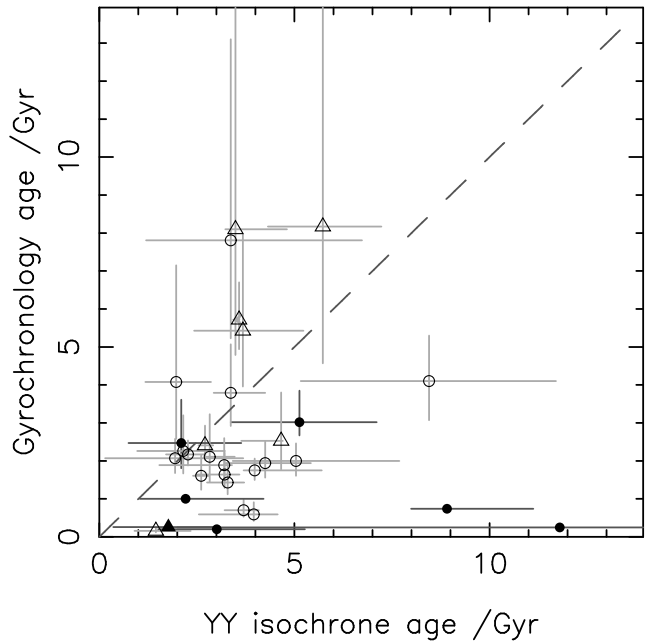}
	\caption{Gyrochronology age as a function of isochrone age for the sub-sample of systems with measured spin--orbit alignment angle, $\lambda$. The dashed line denotes ${\rm age}_{\rm gyro}={\rm age}_{\rm iso}$, and the maximum age on both axes is set to the age of the Universe. Circles mark systems which are judged to be `aligned' according to the criterion of \citet{2010ApJ...718L.145W} and triangles mark `misaligned' systems. As with previous plots, closed symbols denote systems with measured rotation periods and open symbols denote systems with derived rotation periods. Even for this reduced sample, there is a slight tendency for isochrone-fitting ages to be older than those from gyrochronology, even when uncertainties on the ages are taken into account.}
	\label{fig:G4vsY_lambda}
\end{figure}

\begin{figure}
	\centering
 	\includegraphics[width=0.48\textwidth]{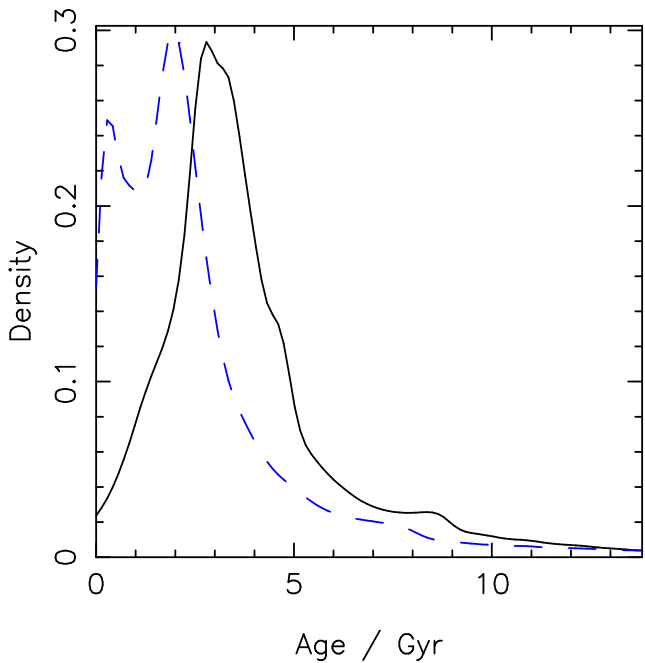}
	\caption{KDEs for the sub-sample of planets with measured spin--orbit alignment angles. The gyrochronology distribution (dashed, blue line) clearly peaks at a younger age than the isochrone fitting distribution (solid, black line), and the median values are similarly offset. However the dual peak of the gyrochronology KDE may be skewing the results towards a smaller offset.}
	\label{fig:G4vsY_dist_lambda}
\end{figure}

Examination of Figure\,\ref{fig:G4vsY_lambda} reveals a similar overall picture to Figure\,\ref{fig:G4vsY}. $52$\,percent of the systems are definitively on the isochrone-fitting side of the ${\rm age}_{\rm gyro}={\rm age}_{\rm iso}$ delineation compared to $13$\,percent on the gyrochronology side, and $58$\,percent of the systems lie within the box bounded by ${\rm age}_{\rm gyro}<4$\,Gyr and ${\rm age}_{\rm iso}<6$\,Gyr. For the full set of systems with measured alignment angles, it therefore seems as though the pattern is similar to that found previously. This is supported by the KDEs (Figure\,\ref{fig:G4vsY_dist_lambda}), with the peak in the gyrochronology distribution appearing to be $\sim0.8$\,Gyr younger than the peak in the isochrone fitting distribution, at $2.0$\,Gyr compared to $2.8$\,Gyr. However, the gyrochronology KDE also displays twin peaks, likely owing to small number statistics, meaning that the true offset could be larger.

A KS test reveals that the probability of a common parent distribution is less than $1$\,percent, and as with the full sample of results, I calculated the $\chi^2$ goodness of fit for this sample to the line ${\rm age}_{\rm Gyro} = {\rm age}_{\rm Iso}$ using equation (\ref{eq:chi2dof}). I found $\chi^2 = 135.1$, $\chi^2_{\rm reduced} = 4.1$, and $P(\chi^2)\sim0$, indicating that, to high significance, the data in this sample are again a poor fit to the hypothesis that the different methods return the same ages. Once again, the systems for which the rotation period has been measured exclusively suggest consistency either with the methods giving similar results, or with older isochrone-fitting ages.

Splitting the sample into `aligned' (circular data) and `misaligned' (triangular data) sets shows that there is little to choose between them. The `aligned' systems appear to show a small bias towards older isochrone-fitting ages, with $61$\,percent of such systems lying to the right of the line denoting equal estimates compared to $9$\,percent lying to the left. The much smaller sample of `misaligned' systems has $25$\,percent of its systems favouring older isochrone-fitting ages and $25$\,percent favouring older gyrochronology ages. The ratio of systems consistent with equal age estimates is $50$\,percent for `aligned' systems and $38$\,percent for `misaligned' systems. There is only one system with both a measured rotation period and a misaligned orbit.

\begin{figure}
	\centering
	\includegraphics[width=0.48\textwidth]{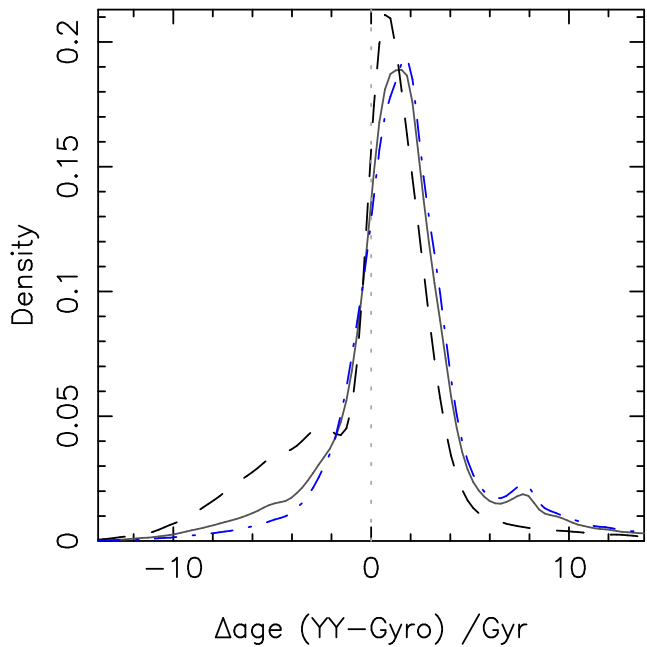}
	\caption{$\Delta{\rm age}$ KDEs for the sub-sample of planets with measured spin--orbit alignment angles (solid, grey distribution), and for the `aligned' (dashed, black distribution) and `misaligned' (dot--dashed, blue distribution) sets. The vertical dotted line marks $\Delta{\rm age}=0$. There is an offset towards positive $\Delta{\rm age}$ for all three distributions, but the effect is weaker for the `aligned' systems.}
	\label{fig:delG4Y_cumdist_lambda}
\end{figure}

Figure\,\ref{fig:delG4Y_cumdist_lambda} displays $\Delta{\rm age}$ KDEs for all of the systems with measured $\lambda$ (grey, solid distribution), and for the `aligned' (black, dashed distribution) and `misaligned' (blue, dot--dashed distribution) sub-samples. Given the preceding discussion, I would expect the peaks of the three KDEs to be broadly similar, which is indeed the case. The distribution for the `misaligned' sub-sample peaks closer to equal ages at $\Delta{\rm age}=0.7$\,Gyr, whilst the aligned distribution peaks at $\Delta{\rm age}\approx1.8$\,Gyr and the overall KDE peaks at $\Delta{\rm age}=1.5$\,Gyr. A 2D KS test on the `aligned' and `misaligned' data returns a probability of $<1$\,percent that they are drawn from the same parent distribution.

For the systems in the `aligned' sample, it is likely that the inclination of the stellar rotation axis to the line of sight, $I_{\rm s}$, is close to $90^{\circ}$ (see the work of \citealt{2010ApJ...719..602S}). However, there is no such guarantee for the `misaligned' systems, and in fact $I_{\rm s}$ may be significantly lower than this value. This would affect the relationship between the measured $v\sin I_{\rm s}$ and the true rotation velocity such that the former would be much smaller than the latter, with the true rotation period therefore being shorter than the value estimated using $v\sin I_{\rm s}$. Since my gyrochronology estimates are based on the derived rotation period in most cases, they would thus be overestimated compared to the actual gyrochronology age; this could be sufficient to bring them in line with the isochrone-fitting estimates. Checking the results of \citet{2010ApJ...719..602S} shows that all of the eight `misaligned' systems are, to varying degrees, rotating more slowly than expected given their age, indicating misalignment of $I_{\rm s}$ and lending support to this idea. This still does not explain why there should be a similar, slightly greater offset for the `aligned' systems however, so it may be that some other mechanism is also acting on the systems concerned.

\begin{figure}
	\centering
	\includegraphics[width=0.48\textwidth]{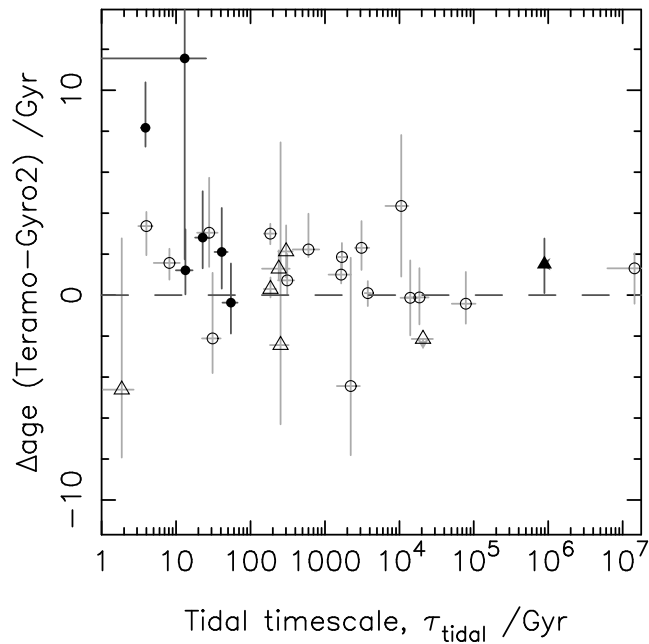}
	\caption{$\Delta{\rm age}$ as a function of tidal time-scale for the sub-sample of planets with measured spin--orbit alignment angles. The `misaligned' systems (triangular data) show no trend, but the `aligned' systems (circular data) hint at a trend for $\Delta{\rm age}$ to increase with decreasing $\tau_{\rm tidal}$. This is based on two data points only though, one of which has substantial $1\sigma$ uncertainties on both quantities.}
	\label{fig:delG4Y-ttide_lambda}
\end{figure}

Looking at $\Delta{\rm age}$ as a function of $\tau_{\rm tide}$ (Figure\,\ref{fig:delG4Y-ttide_lambda}), the small number of `misaligned' systems show no discernible trend with tidal time-scale, as half of them are clustered together at tidal time-scale of between $100$ and $500$\,Gyr. The `aligned' sample shows essentially the same pattern as Figure\,\ref{fig:delG4Y-ttide}, with WASP-19 again being an outlier. The possibility of a trend is again countered by the systems with negative $\Delta{\rm age}$, and postulating anything on the basis of a single datum (the aforementioned WASP-19) would be over interpreting the data.

\section{Discussion}
\label{sec:age_discussion}
At the beginning of Section\,\ref{sec:results} I stated my null hypothesis -- isochronal fitting and gyrochronology are equally accurate, and will produce stellar ages that agree over a large sample. I have demonstrated that this is not quite true, but what is the source of the small disagreement that I have found?\footnote{As I noted in Section \ref{sec:results}, the results presented in this work represent merely one combination of isochrone and gyrochronology ages. My analysis covers five different sets of isochrones and four different gyrochronology formulations, and all $20$ combinations show results that are broadly consistent with those that I have detailed in this work.}

Is isochrone fitting overestimating the ages of the stars in my sample? There does exist a known bias towards older ages when using isochronal analysis, owing to the uneven spacing of data in isochrones near the ZAMS \citep{2010ARAA..48..581S}. \citet{2007ApJ...669.1167B} compared their new gyrochronology ages to isochronal ages for $26$ stars in common between their sample and that of \citet{2007ApJS..168..297T}, finding no correlation between the two. They did however find that the median isochrone age was a factor of $2.7$ higher than the median gyrochronology age, an effect that is substantially greater than the factor of $1.6$ difference in median age that I find.

The YY isochrones are widely used and well studied, but it is possible that any overestimation of ages is a problem with this particular choice of isochrones. It is for this reason that I considered five sets of isochrones, as noted in Section\,\ref{sec:isoimp}. I report the ages obtained using all five model sets in Table\,\ref{tab:isochrones}, and note once more that consideration of the results for any of them reveals similar global patterns to those discussed herein, although the scale of the effect varies. Assuming, based on this, that the discrepancies are not produced by the YY models, then is it a problem with the uncertainties? It is possible that I have underestimated the systematic contributions, which would increase the overlap between the two age estimation methods, but the magnitude of the systematic effects found by \citet{2010MNRAS.408.1689S,2012MNRAS.426.1291S} is in many cases small compared to the uncertainties that I have already derived. However, there is one substantial source of uncertainty that I have neglected during this work. As I stated in Section\,\ref{sec:isoimp}, I neglect the uncertainty in metallicity when calculating the age. Including this factor would increase the uncertainties on my isochrone ages by up to $50$\,percent, and could potentially account for the small discrepancy that I see between the two methods. Finally, is the Delaunay triangulation method that I have implemented producing reliable, consistent age estimates? The overlap between my sample and other studies is insufficient for a comprehensive comparison, but Figure\,\ref{fig:JKTCompare} suggests that the method is working well -- further investigation is required though.

A related possibility is that the stellar parameters I am using are poorly determined. For the stellar density, this is unlikely, as planetary transits allow the density of the host star to be obtained directly from the light curve. However, $T_{\rm eff}$ is usually determined from stellar spectra, but for transiting planet discoveries said spectra are not necessarily of very high resolution, leading to potential inaccuracy and imprecision in the temperature determinations.

The second potential explanation for the discrepancy is that gyrochronology is underestimating the ages of the stars in my sample. The study of $147$ stars with planets by \citet{2010MNRAS.408.1770A}, with a comparison sample of $85$ stars without detected planets, found that stars with planets tended to have greater angular momentum at a given mass than stars without planets. The difference was most pronounced in stars with $M_{\rm s}>1.25M_{\odot}$, and the stars with the most massive planets were found to have the greatest angular momenta relative to the Sun. This would seem to suggest that angular momentum exchange as a result of tidal action could be responsible for the discrepancy in age results. The increase in the angular momentum of a star with one or more planets would in turn decrease its rotation period compared to a star without planets, throwing off the gyrochronology calibration which is carried out using stellar cluster members which have no known companions. But this is at odds with my own findings in Section\,\ref{sec:age_tides}, where there seemed to be no correlation between the tidal time-scale and the difference between the methods.

Another finding of \citet{2010MNRAS.408.1770A} was that stars with planets definitely follow the established relation between rotation and mass that was described by \citet{1967ApJ...150..551K}. Furthermore, they carried out KS tests on the $v\sin I$ distributions of their two samples (stars with and without planets), finding that the results were inconsistent with different origins.

It is important here to again consider the rotation periods that I have used. I have assumed throughout this work that the derived rotation periods are generally reliable, based on the analysis in Section\,\ref{sec:gyroimp}. Assuming that the derived periods are all incorrect leaves an insufficiently large sample for firm conclusions to be drawn, particularly given the magnitude of the uncertainties in some cases. In addition, the gyrochronology relations that I have used may not be calibrated very well for the ages of the stars that I am using. Gyrochronology is generally calibrated using young, open stellar cluster data owing to the large samples of stars with the same age that such data sets provide. However, this is a very different region of parameter space to that occupied by the majority of exoplanet host stars, which tend to be field stars and older in age (as demonstrated by my results). There is therefore no guarantee that the same gyrochronology equations will be valid; efforts to recalibrate gyrochronology for exoplanet hosts using \textit{Kepler} data are ongoing (Angus et al., in preparation), the results of which could have strong implications for this work.

A third possibility is that both methods are inaccurate to some degree, and that there are cases for which both can be considered to be the better option. Although extreme systems such as WASP-19 might be undergoing spin-up that is leading to an underestimation of their gyrochronology age, they seem to be in the minority. I also note that even without tidal spin-up it was impossible to replicate the isochronal age of WASP-19 using gyrochronology. On the other hand, the planets in stellar clusters that I considered demonstrate problems with both age estimation methods.

\citet{2005AA...443..609S} conducted a study of exoplanet host star ages with similar motivation to this work. They focused on estimating age through the use of the chromospheric activity indicator, $R'_{\rm HK}$, but also compared their results to the age as calculated using isochrone fitting, lithium abundance, metallicity, and kinematics. Using a sample of over 100 systems, they found that isochrone ages tended to be older than chromospheric ages, both for their exoplanet host sample and a sample of solar-neighbourhood stars, regardless of which calibration was used for the chromospheric results. They caution though that the dispersions on the two distributions are such that the difference could be nullified. This provides an interesting comparison to the work presented herein.

Chromospheric activity is known to be correlated with stellar rotation \citep[e.g.][]{1963ApJ...138..832W,1972ApJ...171..565S}, so a similar pattern should be expected when comparing chromospheric ages to isochrones as when looking at gyrochronology and isochrones. Unfortunately, \citet{2005AA...443..609S} provide no suggestion for the source of the discrepancy, merely pointing out that the characteristics of the various methods that they use inherently limit them to certain age ranges. But the broad similarity between my results and those of \citet{2005AA...443..609S} is encouraging, even if it provides little additional evidence as to which of the age estimation methods is performing poorly.

Extending my analysis to compare chromospheric ages with the methods that I have already considered would seem an obvious next step for my investigation of exoplanet host star ages, but chromospheric activity is often poorly dealt with by exoplanet studies. The number of planet-hosting stars for which measurements of $\log R'_{\rm hk}$ are available is substantially fewer than the number of planetary systems. In many cases either only a qualitative description is given, or no mention is made of activity in the star. The values that are available are often derived only from single observations, or from observations covering only a short time span (such as the duration of a transit). Given that chromospheric emission often varies periodically, and can do so by significant but unknown factors, makes assessment of stellar age using these data quite inaccurate. There have been studies specifically looking at this metric \citep[e.g.][]{2010ApJ...720.1569K}, but more work is needed.

In future it may be possible to reconcile the differences between gyrochronology and isochrone fitting with additional data. Additional measurements of stellar rotation periods for planet-hosting stars will allow improved gyrochronology estimates by avoiding the systematic errors that are introduced through the use of derived periods, while more precise measurements of stellar parameters such as mass, radius, density, and effective temperature will produce improvements in the results from isochrone fitting. Asteroseismology could also directly improve our age estimates, particularly for older stars for which isochrone fitting can struggle, but will require extensive telescope and analysis time, and relies on the same isochrones as isochrone fitting \citep{2010ARAA..48..581S}.

\section{Conclusion}
\label{sec:ages_conclusion}
I have examined two methods for estimating the ages of exoplanet host stars: isochrone fitting, and gyrochronology. Using a sample of planet-hosting stars, I have shown that there seems to be a small, global discrepancy between the results that are produced by the two methods. This may be linked to stellar effective temperature, with isochrone fitting acting to overestimate the age of the stars in my sample. Examination of planetary systems in stellar clusters, or of the planetary systems for which the stellar rotation period has been measured, suggests that this might be the case, but for the broader sample the possibility that it is a selection effect cannot be ruled out. I investigate the alternative possibility that any discrepancy could be a consequence of tidal interactions affecting the spin-down of planet-hosting stars, finding that the evidence is inconclusive on a sample-wide scale, but that for individual systems tides might play a role. Examining the same possibilities in the context of a sample of systems with measured spin--orbit alignment angles reveals similar results for both `aligned' systems and `misaligned' systems, neither of which show strong evidence for one or the other of the age estimation methods being the cause of the discrepancy.

While the conclusions that I have drawn are potentially interesting, they are limited by the quality and quantity of the available data. The significant uncertainty on many of the derived ages limits the conclusions that can be drawn, while the small sample sizes of the `aligned' and `misaligned' samples is similarly limiting.

\section*{Acknowledgements}
\label{sec:acknowledge}
I would like to thank Andrew Collier Cameron for his comprehensive feedback on an early draft of this work. Other useful feedback came from my PhD examiners Ignas Snellen and Aleks Scholz, and from Ian Bonnell, Rim Fares, and John MacLachlan at the University of St Andrews who gave helpful suggestions for directions in which to take the work. Finally I would like to also thank the referee, who provided extensive comments and constructive pointers for ways in which to refine and focus the manuscript, greatly improving its quality.

This research has made use of: NASA's Astrophysics Data System Bibliographic Services; the SIMBAD data base, operated at CDS, Strasbourg, France; the ArXiv preprint service hosted by Cornell University; the AAVSO Photometric All-Sky Survey (APASS), funded by the Robert Martin Ayers Sciences Fund, and Ren{\'e} Heller's Holt--Rossiter--McLaughlin Encyclopaedia (www.physics.mcmaster.ca/$\sim$rheller/).

Figures\,\ref{fig:ages_DT}-\ref{fig:centroid} are modified from figures available under the GNU Free Documentation License.

\bibliographystyle{mn2e}

\bsp

\appendix
\newpage

\section{Isochronal fitting using Delaunay triangulation}
\label{sec:ages_triangulation}
Delaunay triangulation is a particular method for creating a triangular mesh for a set of data points. It is built upon work by \citet{delaunay1934}, but has since been heavily developed \citep[e.g.][]{shewchuk1996,2006PASP..118.1474P}. I have used the implementation of J. Bernal (see \citealt[for example ]{bernal1988,bernal1991}). 

\begin{figure}
	\centering
	\includegraphics[width=0.48\textwidth]{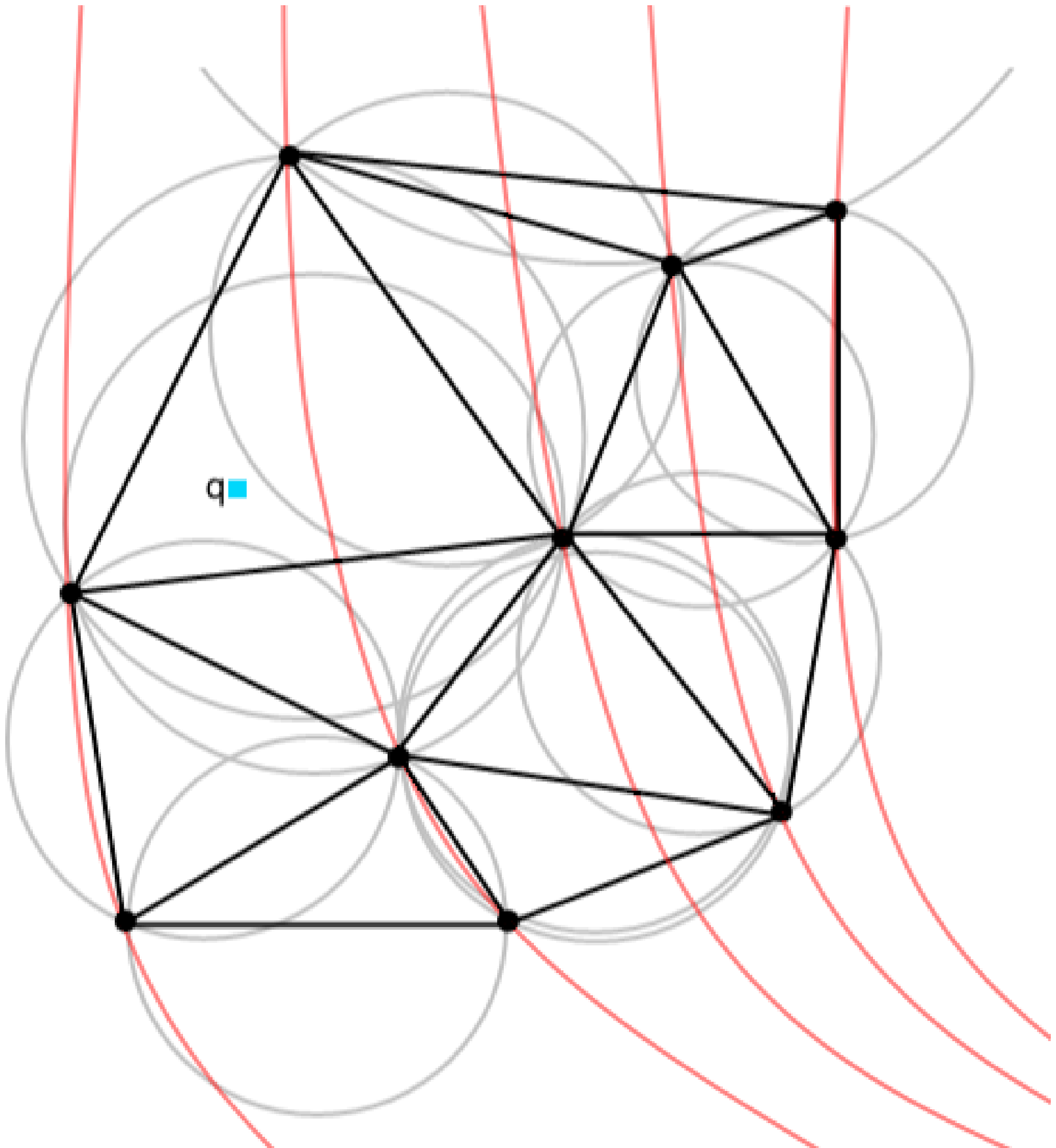}
	\caption{A schematic example of Delaunay triangulation as applied to stellar isochrones. The black circles represent the model data that make up the isochrones (red lines). The blue square, point $\mathbf{q}$, represents the measured stellar data. The triangulation is computed such that the minimum angle across all of the triangles produced is as large as possible. The grey arcs show the circumcircles of the triangles; each circumcircle contains only the data that form the vertices of the corresponding triangle. Once this triangulation is complete, the triangle containing point $\mathit{\mathbf{q}}$ is identified. The vertices of this triangle are then used to interpolate the measured stellar data (see Figure\,\ref{fig:centroid}).}
	\label{fig:ages_DT}
\end{figure}

\begin{figure}
	\centering
	\begin{subfigure}{0.15\textwidth}
		\centering
		\includegraphics[width=\textwidth]{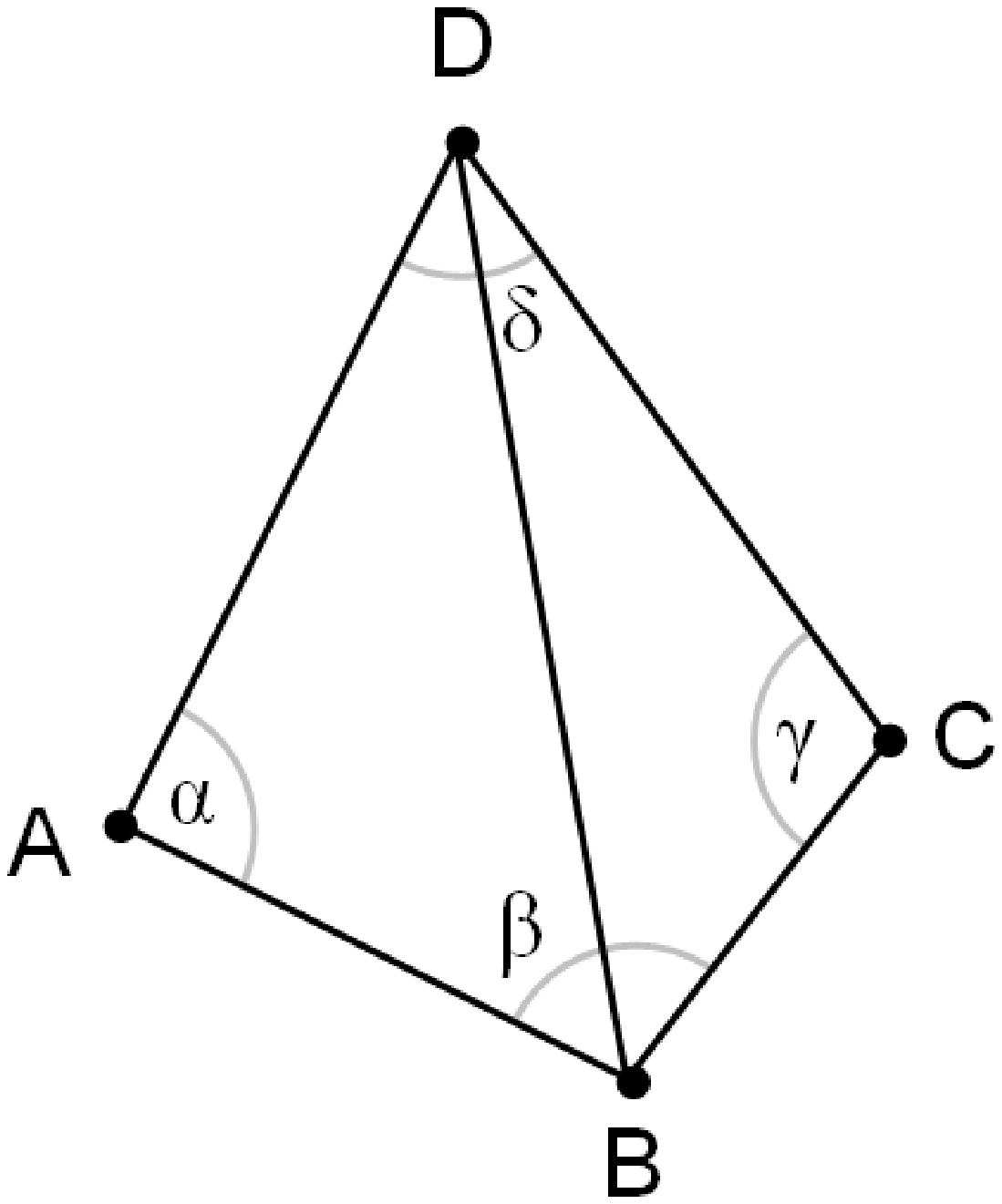}
	\end{subfigure}
	\begin{subfigure}{0.15\textwidth}
		\centering
		\includegraphics[width=\textwidth]{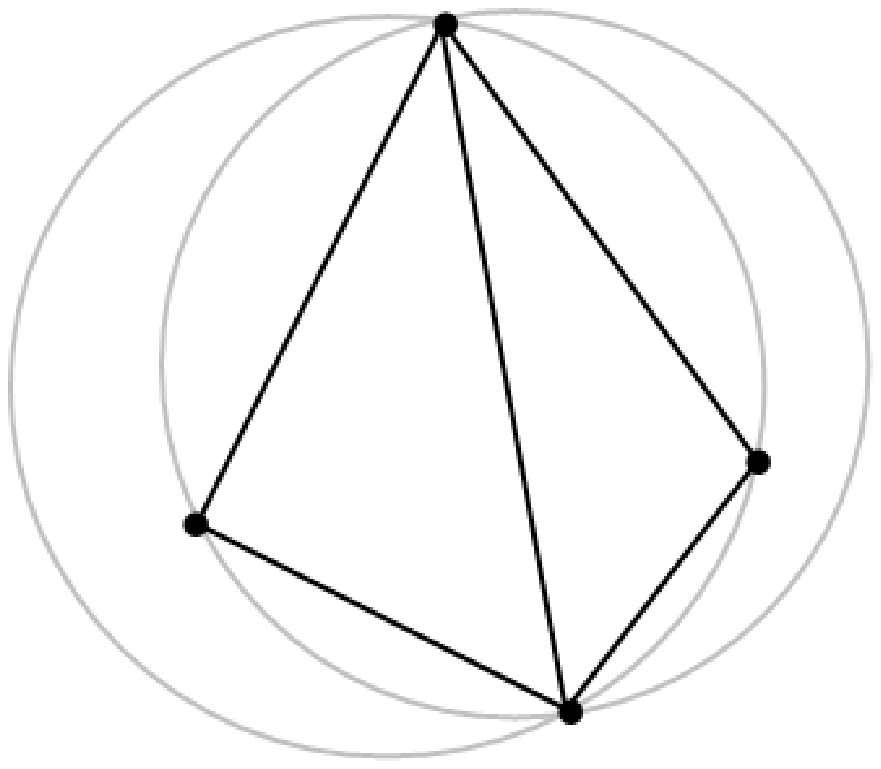}
	\end{subfigure}
	\begin{subfigure}{0.15\textwidth}
		\centering
		\includegraphics[width=\textwidth]{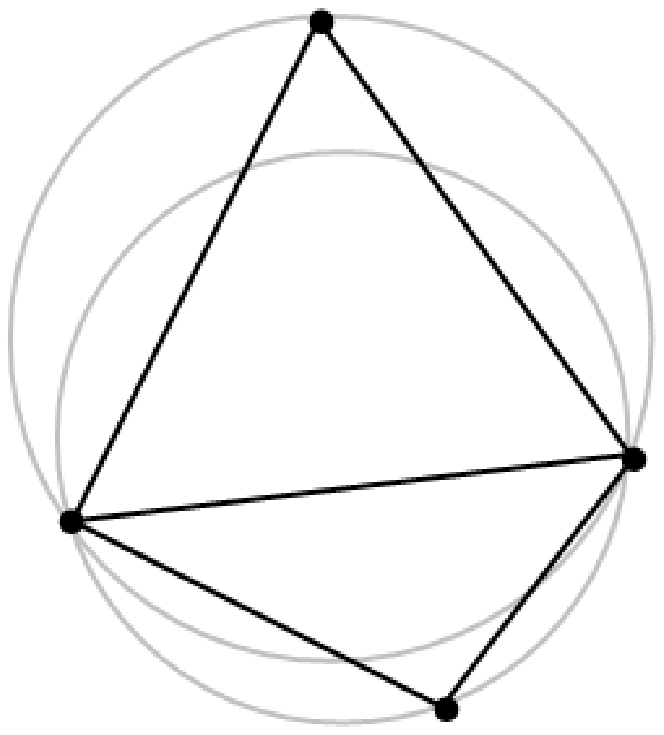}
	\end{subfigure}
	\caption{An example of the edge swapping procedure used  to check for Delaunay compliance, and to optimize the final triangulation. Left: the sum of angles $\alpha$ and $\gamma$ is greater than $180^\circ$. This pair of triangles is therefore not a Delaunay pair. Middle: the circumcircles of the two triangles intersect with the fourth vertex in the pair, also rendering the triangulation non-Delaunay. Right: swapping the line D-B to the line A-C makes this pair of triangles Delaunay compliant. The opposing angles now add up to less than $180^\circ$, and the two circumcircles contain only the vertices of their respective triangles.}
	\label{fig:ages_DTswap}
\end{figure}

There are several specific properties of a Delaunay triangulation that distinguish it from other triangulation methods (see Figure\,\ref{fig:ages_DT}). First, it avoids distorted triangles by maximizing the minimum angle within the triangulation. Secondly, no data other than the vertices of a given triangle may lie within its circumcircle. Thirdly, for any pair of triangles, the sum of the angles opposite to their common side must be less than $180^{\circ}$. If a pair of triangles does not fulfil this third criterion, then swapping the common side such that it bisects those angles creates a Delaunay pair (Figure\,\ref{fig:ages_DTswap}). As each datum is added to the triangulation, the new triangles that are created are checked for Delaunay compliance, and modified if necessary using this procedure.

\subsection{Calculating age}
\label{sec:ages_isocalc}
Once the triangulation is complete, the task of interpolating the measured stellar data is simplified. I identify the component of the triangulation that encloses the measured parameters, and linearly interpolate through the selected triangle using the centroid-based method of \citet{2007nr.book.....P} to identify the age that would be associated with a model datum at the same location as the measured parameters.

\begin{figure}
	\centering
	\includegraphics[width=0.48\textwidth]{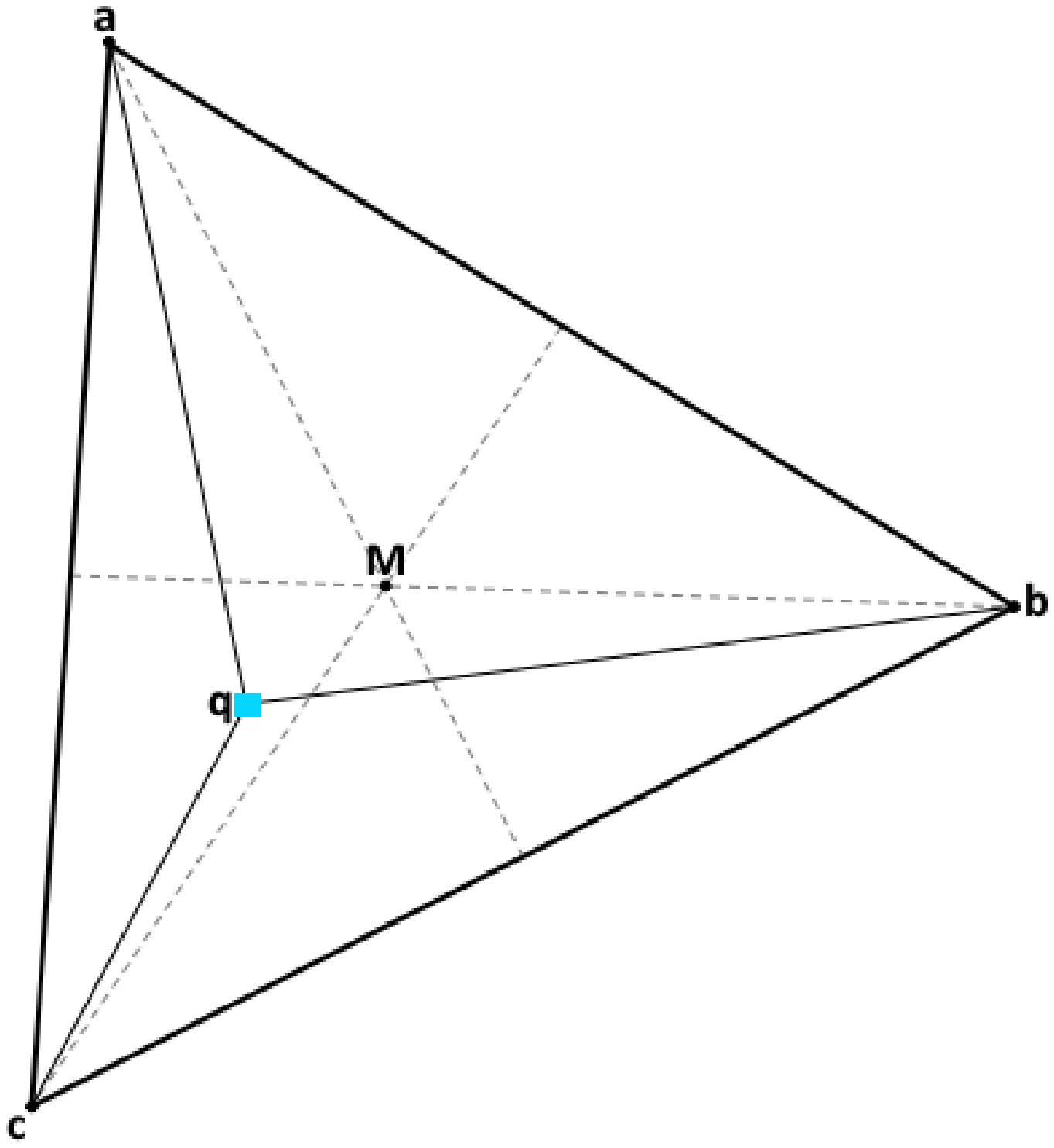}
	\caption{An illustration of the coordinates used for my age interpolation routine. $\mathit{\mathbf{a}}$, $\mathit{\mathbf{b}}$, and $\mathit{\mathbf{c}}$, the black circles, are the vertices of the triangle that has been selected from Figure\,\ref{fig:ages_DT} as containing the measured stellar parameters, which are found at point $\mathit{\mathbf{q}}$, the blue square. $\mathit{\mathbf{M}}$ is the `centroid' of the selected triangle. Each vertex is given a weight according to the ratio of the areas of the component triangles ($\mathit{\mathbf{abq}}$, $\mathit{\mathbf{bcq}}$, and $\mathit{\mathbf{caq}}$) to the area of the enclosing triangle ($\mathit{\mathbf{abc}}$). These weights are then used to interpolate the age at $\mathit{\mathbf{q}}$ according to equation (\ref{eq:datum}).}
	\label{fig:centroid}
\end{figure}

The `centroid' of a triangle lies at the intersection of the lines joining the triangles vertices to the midpoints of their opposing sides (see Figure\,\ref{fig:centroid}). By definition, it is the point where the areas $\mathcal{A}(\mathit{\bf{abM}})$, $\mathcal{A}(\mathit{\bf{bcM}})$, and $\mathcal{A}(\mathit{\bf{caM}})$ are equal, and it's coordinates are given by

\begin{equation}
	M_{i=0,1}=\frac{1}{3}(a_i+b_i+c_i).
	\label{eq:centroid}
\end{equation} By extension, any point in the plane defined by the triangles vertices can be defined as a linear combination of these vertices, with coefficients that sum to unity:

\begin{equation}
	\mathit{\bf{q}}=\alpha \mathit{\bf{a}}+\beta \mathit{\bf{b}}+\gamma \mathit{\bf{c}}
	\label{eq:datum}
\end{equation} For any given point, the coefficients (weights) can be determined using the areas of the plane and of the three component triangles:

\begin{eqnarray}
	\alpha&=&\mathcal{A}(\mathit{\bf{bcq}})/\mathcal{A}(\mathit{\bf{abc}}) \\
	\beta&=&\mathcal{A}(\mathit{\bf{caq}})/\mathcal{A}(\mathit{\bf{abc}}) \\
	\gamma&=&\mathcal{A}(\mathit{\bf{abq}})/\mathcal{A}(\mathit{\bf{abc}})
	\label{eq:interpolate}
\end{eqnarray} Since the [$T_{\rm eff}, (\rho_{\rm s}/\rho_\odot)^{-1/3}$] coordinates for the vertices of the triangle enclosing the measured parameters are known, it is trivial to calculate these weights. The ages known to correspond to the same vertices can then be used alongside the weights to calculate the age corresponding to the measured parameters using equation (\ref{eq:datum}). This method provides a unique solution, as the three vertices of the triangle define a unique plane in three dimensions \citep{2007nr.book.....P}.

The specific property of the Delaunay triangulation to maximize the minimum angle of all triangles is particularly important in this context, as the isochrone data are not distributed uniformly in [$T_{\rm eff}, (\rho_{\rm s}/\rho_\odot)^{-1/3}$] parameter space. Making the triangles as equiangular as possible helps with the interpolation process, as it decreases the chance that two vertices will share an age.

Uncertainties in the calculated age are determined by following the same interpolation procedure using data corresponding to eight points around the error ellipse. These are the extremes of the error bars on $T_{\rm eff}$ and $(\rho_{\rm s}/\rho_\odot)^{-1/3}$, and the points at $45^\circ$ between the error bars. The shape of the isochrones and evolutionary tracks is such that simply using the error bars can underestimate the uncertainty in the age; using the intermediate points helps to alleviate this.

Stellar effective temperatures for the sample were taken from references containing the most recent spectroscopic analyses. Stellar densities were taken from the most recent analyses of the relevant planetary systems (at time of writing); directly listed values were used preferentially, otherwise the density was calculated using the stellar mass and radius. References for these data are given in Table\,\ref{tab:data}.

\section{Age determinations}
\label{sec:tables}

\onecolumn
\begingroup
\begin{longtable}{llllllllll}
\caption{All age estimates for the sample of stars studied herein. Five different sets of isochrone ages, and four different sets of gyrochronology ages, are provided.} \\
\label{tab:isochrones} \\
\hline
System			& \multicolumn{5}{l}{Isochrone age (Gyr)}			& \multicolumn{4}{l}{Gyrochronology age (Gyr)}							\\ [2pt]
				& Padova	& YY	& Teramo	& VRSS	& DSED	& ${\rm age}_1$	& ${\rm age}_2$	& ${\rm age}_3$	& ${\rm age}_4$	\\ [2pt]
\hline
\endfirsthead
\multicolumn{10}{l}{\tablename\ \thetable\ -- \textit{Continued from previous page}} \\
\hline
System			& \multicolumn{5}{l}{Isochrone age (Gyr)}			& \multicolumn{4}{l}{Gyrochronology age (Gyr)}							\\ [2pt]
				& Padova	& YY	& Teramo	& VRSS	& DSED	& ${\rm age}_1$	& ${\rm age}_2$	& ${\rm age}_3$	& ${\rm age}_4$	\\ [2pt]
\hline
\endhead
\hline
\multicolumn{10}{r}{\textit{Continued on next page}} \\
\endfoot
\hline
\endlastfoot
	WASP-1	& $2.75^{+0.27}_{-0.21}$	& $2.71^{+0.21}_{-0.17}$	& $3.13^{+0.78}_{-0.38}$	& $2.82^{+0.36}_{-0.18}$	& $3.22^{+0.25}_{-0.22}$		& $1.42^{+0.24}_{-0.20}$	& $1.61^{+0.23}_{-0.19}$	& $1.57^{+0.24}_{-0.19}$	& $2.41^{+0.52}_{-0.38}$	\\ [2pt]
	WASP-2	& $10.83^{+1.76}_{-1.76}$ & $9.73^{+2.61}_{-2.45}$ & $>8.88$			& $>11.58$			& $10.53^{+2.65}_{-1.98}$	& $1.80^{+1.92}_{-0.65}$	& $8.25^{+8.26}_{-2.94}$	& $7.47^{+7.47}_{-2.67}$	& $7.96^{+9.24}_{-2.97}$	\\ [2pt]
	WASP-4	& $6.73^{+2.93}_{-4.30}$	& $5.13^{+1.98}_{-1.76}$	& $8.23^{+2.76}_{-3.35}$	& $7.53^{+4.26}_{-2.76}$	& $7.43^{+2.00}_{-2.27}$		& $2.17^{+0.57}_{-0.28}$	& $3.36^{+0.83}_{-0.39}$	& $3.14^{+0.77}_{-0.39}$	& $3.02^{+0.83}_{-0.35}$	\\ [2pt]
	WASP-5	& $6.17^{+2.86}_{-2.13}$	& $5.04^{+2.65}_{-1.62}$ & $9.31^{+3.86}_{-1.95}$	& $6.57^{+3.84}_{-1.69}$	& $7.47^{+2.38}_{-1.78}$		& $1.77^{+ 0.46}_{-0.35}$	& $2.35^{+0.50}_{-0.43}$	& $2.25^{+0.49}_{-0.42}$	& $2.00^{+0.46}_{-0.39}$	\\ [2pt]
	WASP-6 	& $>7.94$				& $8.45^{+3.25}_{-3.29}$	& $>11.95$			& $10.87^{+5.63}_{-4.10}$ & $10.40^{+2.92}_{-2.98}$	& $2.53^{+0.69}_{-0.64}$	& $4.53^{+1.23}_{-1.12}$	& $4.20^{+1.17}_{-1.04}$	& $4.10^{+1.20}_{-1.03}$	\\ [2pt]
	WASP-8	& $-$				& $<3.58$				& $<7.10$				& $<6.43$				& $3.19^{+2.66}_{-2.40}$		& $7.22^{+4.36}_{-2.07}$	& $5.86^{+0.97}_{-0.79}$	& $5.48^{+0.94}_{-0.78}$	& $5.72^{+0.98}_{-0.78}$	\\ [2pt]
	WASP-12	& $3.32^{+0.54}_{-0.44}$	& $3.49^{+1.32}_{-0.26}$	& $4.66^{+5.73}_{-0.98}$	& $4.20^{+0.15}_{-1.38}$	& $5.00^{+2.73}_{-0.62}$		& $6.54^{+11.30}_{-3.43}$ & $5.27^{+3.91}_{-1.85}$ & $5.17^{+3.87}_{-1.81}$ & $8.10^{+7.26}_{-3.31}$ \\ [2pt]
	WASP-16	& $4.50^{+4.12}_{-4.25}$	& $3.37^{+3.36}_{-2.17}$ & $5.73^{+4.61}_{-4.01}$	& $5.22^{+4.58}_{-3.53}$	& $5.86^{+3.11}_{-3.10}$		& $5.59^{+3.57}_{-1.82}$	& $7.42^{+4.22}_{-2.23}$	& $6.99^{+4.02}_{-2.11}$	& $7.81^{+5.29}_{-2.58}$	\\ [2pt]
	WASP-19	& $>8.25$				& $8.91^{+2.21}_{-0.92}$	& $>10.51$			& $>9.73$				& $11.37^{+2.79}_{-2.31}$	& $0.53^{+0.21}_{-0.12}$	& $0.86^{+0.07}_{-0.07}$	& $0.80^{+0.08}_{-0.07}$	& $0.74^{+0.05}_{-0.04}$	\\ [2pt]
	WASP-20	& $-$				& $-$				& $-$				& $-$				& $<1.49$					& $0.66^{+0.78}_{-0.37}$	& $0.72^{+0.55}_{-0.37}$	& $0.70^{+0.54}_{-0.36}$	& $0.75^{+0.68}_{-0.40}$	\\ [2pt]
	WASP-21	& $-$				& $12.37^{+2.77}_{-1.90}$ & $15.69^{+0.31}_{-3.46}$ & $13.02^{+3.55}_{-2.06}$ & $13.06^{+1.94}_{-1.97}$	& $27.01^{+64.65}_{-15.36}$ & $8.87^{+12.57}_{-4.07}$ & $8.43^{+11.95}_{-3.86}$ & $9.80^{+16.59}_{-4.85}$ \\ [2pt]
	WASP-22	& $4.58^{+1.73}_{-1.13}$	& $4.25^{+1.17}_{-1.01}$	& $6.27^{+2.03}_{-1.63}$	& $5.23^{+1.96}_{-1.19}$	& $5.81^{+1.28}_{-0.98}$		& $1.99^{+0.73}_{-0.47}$	& $1.88^{+0.36}_{-0.28}$	& $1.82^{+0.36}_{-0.28}$	& $1.94^{+0.57}_{-0.38}$	\\ [2pt]
	WASP-25	& $<3.10$				& $1.94^{+1.75}_{-1.79}$	& $6.09^{+2.77}_{-3.22}$	& $1.18^{+4.34}_{-0.51}$	& $3.25^{+1.79}_{-1.53}$		& $1.35^{+0.33}_{-0.25}$ & $1.91^{+0.42}_{-0.32}$	& $1.78^{+0.40}_{-0.30}$	& $2.07^{+0.51}_{-0.39}$	\\ [2pt]
	WASP-26	& $6.58^{+1.81}_{-1.74}$ & $5.73^{+1.50}_{-1.41}$	& $7.22^{+1.78}_{-1.58}$	& $6.29^{+1.69}_{-1.24}$	& $7.25^{+1.60}_{-1.08}$		& $6.93^{+9.06}_{-3.31}$	& $6.95^{+7.09}_{-2.76}$	& $6.71^{+6.79}_{-2.69}$	& $8.17^{+9.78}_{-3.60}$	\\ [2pt]
	WASP-28	& $2.88^{+3.44}_{-1.34}$	& $1.68^{+2.65}_{-0.96}$	& $2.26^{+2.45}_{-1.84}$	& $1.78^{+2.08}_{-1.68}$	& $3.63^{+2.02}_{-1.94}$		& $1.9^{+ 0.77}_{-0.53}$	& $1.52^{+0.54}_{-0.37}$	& $1.45^{+0.53}_{-0.36}$	& $2.83^{+2.53}_{-1.13}$	\\ [2pt]
	WASP-30	& $2.69^{+0.37}_{-0.17}$	& $2.70^{+0.36}_{-0.24}$	& $3.52^{+0.32}_{-0.60}$	& $2.50^{+0.24}_{-0.49}$	& $3.64^{+0.48}_{-0.43}$		& $1.03^{+0.47}_{-0.25}$	& $0.40^{+0.05}_{-0.04}$	& $0.39^{+0.06}_{-0.05}$	& $0.62^{+0.13}_{-0.10}$	\\ [2pt]
	WASP-32	& $1.43^{+3.52}_{-0.17}$	& $2.10^{+1.54}_{-1.35}$	& $5.22^{+1.59}_{-1.54}$	& $1.67^{+1.95}_{-0.90}$	& $4.48^{+0.98}_{-2.52}$		& $2.07^{+2.53}_{-0.85}$	& $1.75^{+0.49}_{-0.33}$	& $1.68^{+0.47}_{-0.32}$	& $2.47^{+1.14}_{-0.66}$	\\ [2pt]
	WASP-34	& $-$				& $-$				& $-$				& $-$				& $<12.53$ 				& $6.67^{+12.00}_{-3.52}$ & $7.68^{+12.75}_{-3.82	}$ & $7.28^{+12.08}_{-3.64}$ & $7.72^{+15.79}_{-4.12}$	\\ [2pt]
	WASP-35	& $4.70^{+1.80}_{-3.20}$	& $2.98^{+2.16}_{-1.75}$	& $6.00^{+2.88}_{-1.85}$	& $3.73^{+1.75}_{-1.33}$	& $3.80^{+2.49}_{-0.49}$		& $2.55^{+2.67}_{-1.05}$	& $1.70^{+0.59}_{-0.45}$	& $1.62^{+0.57}_{-0.43}$	& $2.13^{+0.88}_{-0.64}$	\\ [2pt]
	WASP-36 	& $2.15^{+2.53}_{-1.96}$	& $1.86^{+1.96}_{-1.24}$	& $2.65^{+2.68}_{-2.16}$	& $<4.01$				& $3.30^{+1.85}_{-1.86}$		& $1.94^{+2.92}_{-0.93}$	& $2.02^{+2.60}_{-0.88}$	& $1.97^{+2.54}_{-0.86}$	& $2.17^{+3.36}_{-1.04}$	\\ [2pt]
	WASP-37	& $>8.32$				& $10.43^{+3.66}_{-3.30}$ & $>8.51$			& $10.69^{+5.49}_{-3.78}$ & $10.31^{+4.01}_{-2.55}$	& $2.79^{+9.68}_{-1.67}$	& $2.89^{+9.04}_{-1.68}$	& $2.72^{+8.52}_{-1.59}$	& $3.16^{+11.80}_{-1.90}$ \\ [2pt]
	WASP-38	& $3.41^{+0.48}_{-0.43}$	& $3.29^{+0.42}_{-0.53}$	& $3.59^{+0.77}_{-0.70}$	& $3.20^{+0.73}_{-0.59}$	& $4.81^{+0.52}_{-0.63}$		& $0.1^{+ 0.26}_{-0.06}$	& $0.94^{+0.13}_{-0.11}$	& $0.92^{+0.15}_{-0.12}$	& $1.43^{+0.54}_{-0.31}$	\\ [2pt]
	WASP-39	& $7.00^{+1.58}_{-5.06}$	& $8.55^{+1.99}_{-4.02}$	& $13.93^{+0.28}_{-3.21}$ & $7.41^{+4.26}_{-4.57}$ & $10.42^{+4.58}_{-1.45}$	& $3.80^{+6.94}_{-1.91}$	& $5.98^{+10.07}_{-2.85}$ & $5.52^{+9.29}_{-2.63}$ & $5.51^{+10.20}_{-2.72}$ \\ [2pt]
	WASP-41	& $>3.01$				& $6.97^{+4.57}_{-3.34}$	& $>3.88$				& $11.10^{+2.63}_{-6.48}$ & $9.07^{+4.85}_{-3.47}$	& $2.95^{+10.23}_{-1.79}$ & $4.70^{+14.72}_{-2.73}$ & $4.40^{+13.73}_{-2.55}$ & $4.07^{+16.48}_{-2.49}$	\\ [2pt]
	WASP-44	& $-$				& $-$				& $2.36^{+0.71}_{-0.71}$	& $<2.65$				& $<2.93$					& $0.77^{+0.74}_{-0.31}$	& $1.59^{+1.31}_{-0.57}$	& $1.52^{+1.26}_{-0.55}$	& $1.36^{+1.20}_{-0.51}$	\\ [2pt]
	WASP-45	& $-$				& $-$				& $0.43^{+4.65}_{-0.01}$	& $<4.80$				& $<3.76$					& $1.13^{+1.14}_{-0.46}$	& $2.73^{+2.57}_{-1.08}$	& $2.52^{+2.38}_{-1.00}$	& $3.36^{+3.59}_{-1.45}$	\\ [2pt]
	WASP-46	& $-$				& $10.84^{+3.81}_{-4.03}$ & $15.52^{+0.48}_{-5.30}$ & $11.50^{+6.50}_{-4.54}$ & $11.44^{+3.56}_{-3.28}$ & $1.73^{+0.74}_{-0.43}$	& $2.19^{+0.32}_{-0.28}$	& $2.09^{+0.33}_{-0.29}$	& $1.74^{+0.30}_{-0.24}$	\\ [2pt]
	WASP-47	& $>10.50$			& $11.28^{+2.94}_{-2.35}$ & $-$				& $>12.62$			& $>11.49$				& $1.69^{+0.86}_{-0.50}$	& $2.65^{+1.30}_{-0.77}$	& $2.47^{+1.22}_{-0.72}$	& $2.13^{+1.15}_{-0.61}$	\\ [2pt]
	WASP-48	& $5.30^{+1.80}_{-1.49}$	& $5.39^{+0.63}_{-1.77}$	& $6.55^{+2.33}_{-0.62}$	& $5.63^{+1.38}_{-1.75}$	& $6.33^{+1.62}_{-0.97}$		& $0.10^{+0.05}_{-0.03}$	& $0.29^{+0.08}_{-0.07}$	& $0.28^{+0.09}_{-0.07}$	& $0.24^{+0.10}_{-0.06}$	\\ [2pt]
	WASP-49	&$7.89^{+4.70}_{-3.70}$	& $6.23^{+2.83}_{-2.33}$	& $9.52^{+4.41}_{-3.55}$	& $7.69^{+4.56}_{-3.29}$	& $7.60^{+2.59}_{-2.54}$		& $13.09^{+15.04}_{-5.79}$ & $17.68^{+18.72}_{-7.29}$ & $16.64^{+17.66}_{-6.86}$ & $23.06^{+31.23}_{-10.51}$	\\ [2pt]
	WASP-50	& $1.06^{+0.80}_{-0.84}$	& $1.86^{+4.41}_{-1.20}$	& $2.21^{+2.79}_{-2.03}$	& $1.13^{+1.15}_{-0.90}$	& $1.28^{+1.85}_{-0.98}$		& $1.01^{+0.18}_{-0.14}$	& $1.87^{+0.17}_{-0.15}$	& $1.74^{+0.17}_{-0.15}$	& $2.23^{+0.50}_{-0.31}$	\\ [2pt]
	WASP-54	& $5.24^{+1.19}_{-1.17}$ & $5.56^{+0.89}_{-0.51}$ & $6.10^{+1.38}_{-0.84}$	& $5.79^{+1.14}_{-0.75}$	& $6.55^{+1.20}_{-0.77}$		& $7.82^{+8.00}_{-3.27}$	& $4.44^{+2.25}_{-1.29}$	& $4.28^{+2.18}_{-1.26}$	& $6.47^{+5.84}_{-2.36}$	\\ [2pt]
	WASP-55	& $6.68^{+2.97}_{-2.11}$	& $5.33^{+2.17}_{-2.35}$	& $8.12^{+2.93}_{-2.82}$	& $4.62^{+3.48}_{-1.87}$	& $6.51^{+2.28}_{-1.68}$		& $2.96^{+4.87}_{-1.51}$	& $2.35^{+2.36}_{-0.92}$	& $2.22^{+2.24}_{-0.87}$	& $2.97^{+3.66}_{-1.34}$	\\ [2pt]
	WASP-57	& $<3.88$				& $2.12^{+1.81}_{-1.81}$	& $3.51^{+3.44}_{-0.94}$	& $<3.26$				& $3.62^{+1.74}_{-1.74}$		& $0.61^{+0.80}_{-0.29}$	& $1.03^{+1.26}_{-0.48}$	& $0.97^{+1.19}_{-0.45}$	& $0.86^{+1.08}_{-0.38}$	\\ [2pt]
	WASP-58	& $11.66^{+0.93}_{-6.12}$ & $9.75^{+3.90}_{-4.66}$ & $11.9^{+ 2.87}_{-4.25}$ & $4.62^{+10.90}_{-0.94}$ & $9.83^{+4.17}_{-0.43}$	& $-$				& $4.18^{+4.43}_{-1.85}$	& $4.02^{+4.26}_{-1.78}$	& $4.54^{+5.97}_{-2.11}$	\\ [2pt]
	WASP-60	& $5.64^{+2.31}_{-2.94}$	& $3.51^{+2.68}_{-1.45}$	& $4.25^{+4.45}_{-1.67}$	& $4.32^{+3.50}_{-1.92}$	& $5.75^{+3.26}_{-1.46}$		& $1.64^{+1.20}_{-0.61}$	& $2.25^{+1.52}_{-0.81}$	& $2.12^{+1.44}_{-0.76}$	& $2.54^{+2.04}_{-0.98}$	\\ [2pt]
	WASP-63	& $8.01^{+1.32}_{-1.21}$ 	& $7.82^{+1.09}_{-1.13}$	& $8.89^{+1.35}_{-1.37}$	& $8.03^{+1.15}_{-1.21}$	& $9.00^{+1.18}_{-1.27}$		& $4.77^{+2.24}_{-1.36}$	& $7.19^{+3.23}_{-1.90}$	& $6.71^{+3.04}_{-1.81}$	& $7.12^{+3.48}_{-2.05}$	\\ [2pt]
	WASP-64	& $>7.89$				& $8.94^{+3.15}_{-2.55}$	& $11.42^{+4.58}_{-4.15}$ & $11.42^{+4.90}_{-3.19}$ & $10.96^{+2.72}_{-2.93}$	& $1.21^{+0.82}_{-0.41}$	& $1.84^{+1.11}_{-0.59}$	& $1.73^{+1.04}_{-0.56}$	& $1.71^{+1.18}_{-0.59}$	\\ [2pt]
	WASP-65	& $>8.26$				& $8.92^{+1.87}_{-1.97}$	& $11.42^{+4.01}_{-2.75}$ & $11.31^{+3.14}_{-2.49}$ & $10.80^{+2.36}_{-2.03}$	& $1.39^{+0.61}_{-0.39}$	& $2.09^{+0.68}_{-0.45}$	& $2.02^{+0.67}_{-0.45}$	& $1.51^{+0.50}_{-0.33}$	\\ [2pt]
	WASP-70A & $8.30^{+1.70}_{-1.90}$	& $4.68^{+3.47}_{-1.31}$	& $8.43^{+3.55}_{-0.64}$	& $8.26^{+0.83}_{-4.44}$	& $9.13^{+1.88}_{-2.94}$		& $4.84^{+3.15}_{-1.64}$	& $7.19^{+4.34}_{-2.37}$	& $6.74^{+4.08}_{-2.26}$	& $8.29^{+5.62}_{-2.89}$\\ [2pt]
	WASP-71	& $3.27^{+0.33}_{-0.74}$	& $3.21^{+0.38}_{-0.74}$	& $3.16^{+0.55}_{-0.46}$	& $3.04^{+0.50}_{-0.26}$	& $3.67^{+0.76}_{-0.30}$		& $1.15^{+1.11}_{-0.41}$	& $1.37^{+0.24}_{-0.20}$	& $1.33^{+0.24}_{-0.21}$	& $1.64^{+0.58}_{-0.35}$	\\ [2pt]
	WASP-75	& $1.02^{+2.26}_{-0.11}$	& $2.08^{+0.60}_{-0.95}$	& $3.92^{+1.67}_{-1.73}$	& $1.48^{+2.32}_{-0.34}$	& $3.91^{+1.26}_{-1.13}$		& $2.22^{+1.42}_{-0.76}$	& $2.11^{+0.96}_{-0.57}$	& $2.06^{+0.95}_{-0.56}$	& $5.92^{+4.43}_{-1.78}$	\\ [2pt]
	WASP-77A & $6.29^{+5.13}_{-3.10}$	& $5.34^{+2.19}_{-2.08}$	& $>7.81$				& $9.48^{+5.41}_{-4.08}$	& $7.82^{+2.75}_{-2.43}$		& $0.63^{+0.09}_{-0.07}$	& $1.27^{+0.17}_{-0.15}$	& $1.21^{+0.18}_{-0.15}$	& $0.92^{+0.12}_{-0.09}$	\\ [2pt]
	WASP-84	& $-$				& $-$				& $-$				& $-$				& $-$					& $0.31^{+0.08}_{-0.06}$	& $0.60^{+0.10}_{-0.08}$	& $0.55^{+0.10}_{-0.08}$	& $0.56^{+0.08}_{-0.06}$	\\ [2pt]
	WASP-95	& $2.90^{+3.07}_{-1.47}$	& $2.56^{+2.18}_{-0.68}$	& $4.89^{+3.46}_{-2.44}$	& $4.87^{+1.83}_{-2.44}$	& $3.91^{+3.27}_{-1.30}$		& $1.62^{+1.10}_{-0.57}$	& $2.76^{+1.35}_{-0.81}$	& $2.62^{+1.30}_{-0.78}$	& $2.90^{+1.76}_{-0.96}$	\\ [2pt]
	WASP-96	& $6.81^{+5.78}_{-2.44}$	& $5.17^{+4.32}_{-1.10}$	& $9.50^{+6.50}_{-2.83}$	& $-$				& $9.80^{+3.99}_{-3.01}$		& $4.08^{+19.96}_{-2.59}$ & $7.30^{+32.62}_{-4.58}$ & $6.98^{+31.16}_{-4.39}$ & $5.59^{+28.77}_{-3.56}$	\\ [2pt]
	WASP-97	& $3.47^{+2.46}_{-2.97}$	& $3.21^{+1.40}_{-1.41}$	& $5.82^{+3.34}_{-2.38}$	& $4.57^{+4.11}_{-2.47}$	& $5.29^{+2.01}_{-3.10}$		& $12.63^{+25.37}_{-6.63}$ & $15.09^{+27.61}_{-7.61}$ & $14.28^{+26.32}_{-7.21}$ & $15.53^{+32.60}_{-8.09}$ \\ [2pt]
	WASP-98	& $<8.09$				& $6.71^{+5.43}_{-3.66}$	& $>4.14$				& $4.74^{+3.47}_{-3.47}$	& $5.68^{+2.87}_{-2.87}$		& $22.43^{+7.27}_{-4.38}$ & $27.11^{+2.74}_{-2.39}$ & $25.43^{+2.85}_{-2.47}$ & $28.84^{+4.40}_{-3.38}$	\\ [2pt]
	WASP-99	& $2.15^{+0.85}_{-0.89}$	& $2.45^{+0.76}_{-0.30}$	& $2.60^{+1.26}_{-1.10}$	& $2.60^{+1.12}_{-0.40}$	& $3.00^{+0.69}_{-0.73}$		& $-$				& $1.56^{+0.31}_{-0.24}$	& $1.49^{+0.31}_{-0.24}$	& $2.68^{+1.32}_{-0.73}$	\\ [2pt]
	CoRot-2	& $2.41^{+4.30}_{-0.95}$	& $3.01^{+2.26}_{-1.50}$	& $4.29^{+3.21}_{-2.59}$	& $4.35^{+3.08}_{-2.42}$	& $5.01^{+1.61}_{-0.91}$		& $0.05^{+0.01}_{-0.01}$	& $0.17^{+0.02}_{-0.02}$	& $0.16^{+0.02}_{-0.02}$	& $0.20^{+0.01}_{-0.01}$	\\ [2pt]
	CoRot-18	& $-$				& $11.80^{+5.71}_{-9.80}$ & $>1.84$			& $>8.49$				& $>7.08$					& $0.10^{+0.09}_{-0.03}$	& $0.26^{+0.04}_{-0.04}$	& $0.25^{+0.04}_{-0.04}$	& $0.25^{+0.03}_{-0.03}$	\\ [2pt]
	CoRot-19	& $3.69^{+1.01}_{- 0.33}$	& $4.66^{+0.04}_{-1.02}$	& $5.06^{+0.63}_{-0.95}$	& $3.46^{+1.93}_{-0.06}$	& $5.85^{+0.49}_{-0.70}$		& $0.53^{+0.31}_{-0.17}$	& $1.26^{+0.49}_{-0.31}$	& $1.16^{+0.45}_{-0.29}$	& $2.53^{+1.27}_{-0.75}$	\\ [2pt]
	HAT-P-1	& $<2.98$				& $2.15^{+1.07}_{-1.18}$	& $1.79^{+1.88}_{-1.39}$	& $2.03^{+1.52}_{-0.68}$	& $2.72^{+0.98}_{-1.32}$		& $0.28^{+0.71}_{-0.16}$	& $2.43^{+0.91}_{-0.59}$	& $2.41^{+0.92}_{-0.59}$	& $2.26^{+0.94}_{-0.59}$	\\ [2pt]
	HAT-P-4 	& $4.36^{+0.67}_{-0.85}$	& $3.98^{+1.72}_{-0.28}$	& $6.14^{+0.90}_{-0.64}$	& $4.74^{+1.72}_{-1.00}$	& $5.20^{+1.78}_{-0.68}$		& $1.40^{+0.29}_{-0.22}$	& $1.82^{+0.28}_{-0.24}$	& $1.75^{+0.28}_{-0.24}$	& $1.75^{+0.31}_{-0.25}$	\\ [2pt]
	HAT-P-8	& $3.64^{+0.53}_{-0.43}$	& $3.70^{+0.39}_{-0.49}$	& $3.26^{+0.35}_{-0.46}$	& $3.64^{+0.33}_{-0.85}$	& $5.00^{+0.43}_{-1.16}$		& $1.45^{+1.80}_{-0.57}$	& $0.47^{+0.09}_{-0.07}$	& $0.46^{+0.09}_{-0.07}$	& $0.70^{+0.28}_{-0.17}$	\\ [2pt]
	HAT-P-13	& $8.40^{+1.48}_{-1.70}$	& $5.83^{+0.51}_{-2.00}$	& $8.98^{+1.50}_{-1.37}$	& $7.64^{+1.44}_{-1.26}$	& $6.50^{+1.97}_{-1.13}$		& $8.29^{+5.76}_{-2.92}$	& $14.94^{+8.96}_{-4.62}$ & $14.29^{+8.52}_{-4.44}$ & $14.17^{+9.84}_{-4.78}$ \\ [2pt]
	HAT-P-16	& $1.39^{+1.06}_{-0.96}$	& $1.97^{+0.89}_{-0.79}$	& $1.80^{+0.83}_{-1.21}$	& $1.76^{+1.13}_{-1.03}$	& $2.50^{+0.82}_{-0.78}$		& $4.19^{+4.46}_{-1.78}$	& $2.49^{+1.31}_{-0.72}$	& $2.42^{+1.29}_{-0.71}$	& $4.08^{+3.07}_{-1.50}$	\\ [2pt]
	HAT-P-23	& $3.94^{+1.74}_{-1.59}$	& $3.96^{+0.61}_{-1.41}$	& $4.57^{+2.06}_{-1.31}$	& $4.65^{+1.77}_{-1.76}$	& $4.88^{+0.86}_{-1.37}$		& $0.06^{+0.14}_{-0.03}$	& $0.66^{+0.34}_{-0.19}$	& $0.64^{+0.33}_{-0.19}$	& $0.59^{+0.32}_{-0.17}$	\\ [2pt]
	HAT-P-32	& $1.12^{+1.10}_{-0.89}$	& $1.45^{+0.89}_{-0.55}$	& $0.96^{+1.36}_{-0.69}$	& $0.94^{+0.96}_{-0.51}$	& $3.08^{+0.73}_{-1.09}$		& $0.17^{+0.24}_{-0.07}$	& $0.14^{+0.02}_{-0.02}$	& $0.14^{+0.03}_{-0.02}$	& $0.16^{+0.07}_{-0.04}$	\\ [2pt]
	HD\,149026	& $2.54^{+0.24}_{-0.23}$	& $2.61^{+0.20}_{-0.21}$	& $2.76^{+0.34}_{-0.26}$	& $2.63^{+0.22}_{-0.29}$	& $3.02^{+0.29}_{-0.21}$	& $-$				& $1.08^{+0.26}_{-0.19}$	& $1.05^{+0.26}_{-0.19}$	& $1.61^{+0.53}_{-0.37}$	\\ [2pt]
	HD\,17156	& $3.23^{+0.75}_{-0.47}$	& $3.37^{+0.88}_{-0.44}$	& $3.38^{+1.16}_{-0.63}$	& $3.75^{+0.44}_{-0.97}$	& $4.00^{+0.29}_{-0.37}$	& $-$				& $2.75^{+0.54}_{-0.45}$	& $2.67^{+0.54}_{-0.45}$	& $3.79^{+1.28}_{-0.87}$	\\ [2pt]
	HD\,209458	& $1.83^{+0.55}_{-0.44}$	& $2.27^{+0.45}_{-0.56}$	& $2.65^{+0.92}_{-0.51}$	& $1.92^{+0.59}_{-0.42}$	& $3.87^{+0.76}_{-0.07}$	& $0.21^{+0.53}_{-0.11}$	& $1.86^{+0.25}_{-0.21}$	& $1.83^{+0.27}_{-0.22}$	& $2.17^{+0.37}_{-0.29}$	\\ [2pt]
	HD\,80606	& $4.56^{+1.73}_{-1.82}$	& $3.68^{+1.55}_{-1.25}$	& $7.83^{+2.21}_{-2.18}$	& $7.34^{+2.46}_{-1.89}$	& $5.73^{+1.64}_{-1.21}$	& $0.91^{+2.28}_{-0.50}$	& $6.18^{+2.68}_{-1.66}$	& $5.84^{+2.56}_{-1.58}$	& $5.43^{+2.53}_{-1.47}$	\\ [2pt]
	Kepler-17	& $3.18^{+2.77}_{-2.96}$	& $2.21^{+2.00}_{-1.17}$	& $4.06^{+4.14}_{-2.18}$	& $5.07^{+2.24}_{-2.38}$	& $4.08^{+2.27}_{-1.65}$		& $0.67^{+0.94}_{-0.27}$	& $1.13^{+0.09}_{-0.08}$	& $1.06^{+0.09}_{-0.08}$	& $1.00^{+0.08}_{-0.07}$	\\ [2pt]
	Kepler-30	& $-$				& $-$				& $<3.14$				& $<3.39$				& $1.76^{+1.40}_{-1.37}$		& $-$				& $1.88^{+0.27}_{-0.22}$	& $1.75^{+0.29}_{-0.23}$	& $1.57^{+0.09}_{-0.09}$	\\ [2pt]
	Kepler-63	& $<2.90$				& $1.77^{+1.25}_{-1.41}$	& $3.78^{+2.09}_{-2.24}$	& $2.77^{+2.10}_{-1.05}$	& $3.42^{+1.02}_{-1.42}$		& $0.16^{+0.05}_{-0.03}$	& $0.28^{+0.02}_{-0.02}$	& $0.26^{+0.02}_{-0.02}$	& $0.26^{+0.004}_{-0.003}$ \\ [2pt]
	KOI-94	& $3.16^{+0.49}_{-1.58}$	& $3.20^{+0.20}_{-1.66}$	& $3.55^{+0.40}_{-0.61}$	& $2.38^{+1.58}_{-0.69}$	& $<5.00$					& $0.69^{+0.31}_{-0.19}$	& $1.20^{+0.29}_{-0.24}$	& $1.17^{+0.30}_{-0.24}$	& $1.89^{+0.71}_{-0.48}$	\\ [2pt]
	TrES-02	& $<4.35$				& $3.15^{+1.40}_{-1.29}$	& $4.10^{+1.87}_{- 2.10}$	& $3.25^{+1.91}_{-2.13}$	& $4.45^{+1.46}_{-1.25}$		& $9.28^{+28.58}_{-5.27}$ & $14.40^{+40.17}_{-7.96}$ & $13.60^{+37.92}_{-7.52}$ & $19.09^{+64.80}_{-11.27}$ 				\\ [2pt]
	TrES-04	& $3.02^{+0.55}_{-0.63}$	& $2.83^{+0.64}_{-0.13}$	& $2.78^{+0.53}_{-0.66}$	& $2.68^{+0.65}_{-0.21}$	& $3.75^{+0.49}_{-0.76}$		& $-$				& $1.35^{+0.44}_{-0.29}$	& $1.34^{+0.45}_{-0.30}$	& $2.11^{+1.12}_{-0.66}$	\\ [2pt]
\end{longtable}
\endgroup
\twocolumn

\label{lastpage}

\end{document}